\documentclass[secnumarabic,amssymb, nobibnotes, aps, prc,twocolumn]{revtex4}

\usepackage{graphicx}
\usepackage{amssymb}
\usepackage{amsmath}
\usepackage{bm}
\usepackage{kotex}
\usepackage{times}
\usepackage{xcolor}
\usepackage{ulem}
\usepackage{cancel}
\usepackage{verbatim}
\usepackage{pifont}
\usepackage{rotating}
\usepackage{subfigure}
\usepackage{multirow}

\usepackage[colorlinks=true,linkcolor=red,citecolor=blue]{hyperref}

\newcommand{\be}{\begin{equation}}
\newcommand{\ee}{\end{equation}}
\newcommand{\bea}{\begin{eqnarray}}
\newcommand{\eea}{\end{eqnarray}}

\newcommand{\etal}{{\it et al.}}

\begin{document}
\setcounter{page}{1}

\title{ Bubble nuclei with shape coexistence in even-even isotopes of Hf to Hg }
\author{Yong-Beom \surname{Choi}}
\email{ybchoi@pusan.ac.kr}
\author{Chang-Hwan \surname{Lee}}
\email{clee@pusan.ac.kr}
\affiliation{Department of Physics, Pusan National University, Busan 46241, Korea}
\author{Myeong-Hwan \surname{Mun}}
\email{aa3101@gmail.com}
\affiliation{Korea Institute of Science and Technology Information, Daejeon 34141, Korea}
\affiliation{Department of Physics and Origin of Matter and Evolution of Galaxy (OMEG) Institute, Soongsil University, Seoul 156-743, Korea}
\author{Youngman \surname{Kim}}
\email{ykim@ibs.re.kr}
\affiliation{Rare Isotope Science Project, Institute for Basic Science, Daejeon 34000, Korea}

\date[]{}

\begin{abstract}
The shape of a nucleus is one of fundamental nuclear properties.  We perform a systematic investigation of bubble nuclei that also exhibit shape coexistence  in  Hf, W, Os, Pt and Hg even-even isotopes using the deformed relativistic Hartree-Bogoliubov theory in continuum. For a systematic study, we first consider nuclear bubble structures and shape coexistence separately. We confirm that deformations and pairing correlations hinder bubble structures by comparing our results with those from relativistic continuum Hartree-Bogoliubov theory that assumes spherical symmetry in nuclei. We then predict candidate isotopes with both bubble structure and shape coexistence. We observe that the depletion fraction factor that characterizes bubble structure is mostly smaller in oblate deformation than in prolate, while some isotopes such as $^{206}$Os have bubble structures both in oblate and prolate deformations. We compare the proton single-particle energy levels for the candidates of shape coexistence both with only prolate bubble structure and with prolate and oblate bubble structures.
\end{abstract}

\pacs{????}


\maketitle


\section{Introduction}

A nucleus, a collection of quantum marbles (nucleons), is by itself a marvelous object showing variety of characters due to quantum nature of its constituents. Understanding how nucleons form a nucleus exhibiting exotic features is unquestionably an important problem and attracts much attention in recent years partly thanks to the new generation of rare isotope beam facilities for production of more exotic nuclei. Eventually, one has to understand how shell and collective properties of a nucleus emerge from more fundamental theories such as quantum chromodynamics.

The shape of a nucleus is one of its most fundamental properties. Most nuclei have a spherical or ellipsoidal shape. However, some of them exhibit exotic features such as pear shape, bubble structure, and shape coexistence. Understanding the exotic features of nuclei provides insight into how neutrons and protons form a nucleus.

The bubble structure in nuclei is a novel exotic nuclear phenomenon which is characterized by a depleted central density. The origin of bubble structure is closely related to a low occupation of the s-state near the Fermi surface. The first experimental evidence of the bubble structure in the unstable nucleus $^{34}$Si was  reported using $\gamma$-ray spectroscopy in Ref. ~\cite{Mutschler:2016ecc}. Electron scattering on unstable nuclei is the most direct way to extract the charge density distribution of a nucleus. Recently, information on the nuclear shape of $^{132}$Xe was extracted  at the self-confining radioactive isotope ion target (SCRIT) facility~\cite{Tsukada:2017llu}. Theoretically, there have been ample studies on the bubble nuclei~\cite{Wilson:1946zza, Wong:1972wum, ToddRutel:2004tu, Khan:2007ji, Grasso:2009zza, Nakada:2012fh, Wu:2013isa, Li:2016cgj, Duguet:2016wwr, Schuetrumpf:2017qeq, Phuc:2018fqo, Wu:2018qaw, Saxena:2018qgd, Li:2018ouf, Kumawat:2020cxl, Arthuis:2020toz}.
Interestingly, the transport model simulations of heavy ion collisions also showed possible bubble structures in nuclear matter \cite{Bauer:1992qv, Li:1993obb, Xu:1993zz, Cherevko:2014csa}. Recently, a method to probe the bubble structure in  heavy ion collisions using the $\pi^+/\pi^-$ ratio was suggested~\cite{Yong:2016zas}.

Shape coexistence is another important exotic nuclear property. A nucleus can exhibit different nuclear shapes with a relatively tiny energy difference compared to the total binding energy. Nuclei that exhibit shape coexistence have several minima in the potential energy surface or curve. For a detailed discussion on shape coexistence we refer to \cite{Wood:1992zza, Gade:2016xiy}, and for recent experimental results we refer to~\cite{Sarazin:1999pe, Olaizola:2019hzn,Garrett:2019hym, Siciliano:2020akh, Garcia:2020oxs, Iskra:2020ces}.

In this work, we seek for an exotic nucleus that exhibits both bubble structure and shape coexistence in Hf, W, Os, Pt, and Hg isotopes. To this end, we work in the framework of the deformed relativistic Hartree-Bogoliubov theory in continuum (DRHBc)~\cite{Zhou:2009sp, Li:2012gv, Meng:2015hta, Zhang:2020wvp}, which includes self-consistently pairing, the continuum and deformation effects, with the PC-PK1 density functional~\cite{Zhao:2010hi}. 
In DRHBc, the axial deformation is considered and the coexistence of spherical, oblate and prolate shapes can be studied.
The effect of pairing and deformation on bubbles can also be studied. However, since the triaxial shapes are not included in DRHBc, the shape coexistence with triaxial shape cannot be investigated in the present work.

The DRHBc theory used in this work has been successfully applied to various interesting phenomena of exotic nuclei; e.g., neutron drip line change with axial deformation~\cite{In:2020asf}, the shape decoupling effect which produces different shapes of core and halo~\cite{Sun:2018ekv, Sun:2020tas, Sun:2021alk}, and the existence of bound nuclei beyond neutron drip line~\cite{Pan:2021oyq, He:2021thz, Zhang:2021ize}.

Even though we focus on the PC-PK1~\cite{Zhao:2010hi},
various relativistic mean-field models
have been used to describe the properties of nuclear ground states and low-lying excited states. The triaxial ground state and shape coexistences in Mo and Ru isotopes \cite{Abusara:2017yzy} have been investigated using DD-ME2~\cite{Lalazissis:2005de} and DD-PC1~\cite{Niksic:2008vp}. Triple-shape coexistence and super deformation of Pb isotopes~\cite{Naz:2018fsj} have been investigated using DD-ME2~\cite{Lalazissis:2005de}, DD-PC1~\cite{Niksic:2008vp}, and NL3*~\cite{Lalazissis:2009zz}. Bubble nuclei~\cite{Shukla:2014bsa} have been studied with NL065~\cite{Reinhard:1989zi}. The low-lying excited states of Mg isotopes \cite{Yao:2010at} and Kr \cite{Yao:2014vba} have been studied with PC-F1 \cite{Burvenich:2001rh} and PC-PK1 \cite{Zhao:2010hi}, respectively. 

We first investigate bubble structures by introducing bubble parameters using the maximum density averaged over the deformed shell. We find bubble nuclei such as $^{256}$Hf, $^{258}$W, $^{260}$Os, $^{262}$Pt, and so on. We compare our results with the ones from relativistic continuum Hartree-Bogoliubov (RCHB) theory with spherical symmetry~\cite{Zhou:2009sp,Meng:1996zz, Meng:1998axq, Xia:2017zka} to discuss the effects of deformations and pairing on bubble structures.

After a brief study of shape coexistence in Hf, W, Os, Pt, and Hg isotopes, we investigate a more exotic nucleus that exhibits both bubble structure and shape coexistence and find some candidates such as $^{202}$Hf, $^{254}$W, and $^{206}$Os.
We also observe that some isotopes such as $^{206}$Os have bubble structure  both in prolate and oblate shapes, while some of them, $^{196}$Os for example, possess it only in prolate shapes. We then discuss the proton single-particle energy levels for $^{206}$Os and $^{196}$Os.

This paper is organized as follows. We first introduce the DRHBc framework  and show the quadrupole deformations which are relevant to the bubble structure and shape coexistence in Section \ref{chap_DRHBc_framework}. We then present our results on bubble nuclei in Section \ref{bubble-C} and those on shape coexistence in Section \ref{scoex}. Isotopes with both bubble structure and shape coexistence are presented in Section \ref{bcsc}. We finally summarize this work in Section \ref{sec5}.


\section{DRHBc Framework} 
\label{chap_DRHBc_framework}


To study nuclear properties we work with the point-coupling model~\cite{Nikolaus:1992zz, Burvenich:2001rh, Zhao:2010hi, A_textbook_of_Jie_Meng} whose Lagrangian density is given by
\bea
{\cal L} &=& \bar{\psi} \left(i\gamma_\mu \partial^\mu - m \right)\psi\,     \nonumber\\
            &-&\frac{1}{2} \, \alpha_{S} \left(\bar{\psi}\psi \right) \left(\bar{\psi}\psi \right)\,
                  -\frac{1}{2} \, \alpha_{V} \left(\bar{\psi} \gamma_\mu \psi \right) \left(\bar{\psi} \gamma^\mu \psi \right)\, \nonumber \\
            &-& \frac{1}{2} \, \alpha_{TV} \left(\bar{\psi} \vec{\tau} \gamma_\mu \psi \right) \left(\bar{\psi} \vec{\tau} \gamma^\mu \psi \right)\, \nonumber\\
            &-&\frac{1}{3}\beta_S \left(\bar{\psi}\psi \right)^3 -\frac{1}{4}\gamma_S \left(\bar{\psi}\psi \right)^4
                   -\frac{1}{4} \, \gamma_{V} \left[\left(\bar{\psi} \gamma_\mu \psi \right) \left(\bar{\psi} \gamma^\mu \psi \right)\right]^2\,    \nonumber\\
            &-&\frac{1}{2} \delta_S \partial_\nu \left(\bar{\psi}\psi \right) \partial^\nu \left(\bar{\psi}\psi \right)
                  -\frac{1}{2} \delta_V \partial_\nu \left(\bar{\psi} \gamma_\mu \psi \right)
                   \partial^\nu \left(\bar{\psi} \gamma^\mu \psi \right) \nonumber \\
            &-&\frac{1}{2} \delta_{TV} \partial_\nu \left(\bar{\psi} \vec{\tau} \gamma_\mu \psi \right)
                  \partial^\nu \left(\bar{\psi} \vec{\tau} \gamma_\mu \psi \right)   \nonumber\\
            &-&\frac{1}{4} F^{\mu \nu} F_{\mu \nu} - e \frac{1-\tau_3}{2} \bar{\psi} \gamma^\mu \psi A_\mu ,
\label{Lag}
\eea
where $m$ is the nucleon mass, $\alpha_{S}$, $\alpha_{V}$ and $\alpha_{TV}$ are the coupling constants for four-fermion contact interactions. The terms with $\beta_{S}$, $\gamma_{S}$ and  $\gamma_{V}$ simulate the effects of density dependence and those with  $\delta_{S}$, $\delta_{V}$ and $\delta_{TV}$  mimic the finite range effects. $A_\mu$ and $F_{\mu \nu}$ are the four-vector potential and field strength tensor of the electromagnetic field, respectively. The subscripts $S, \, V$ and $TV$ are the abbreviation for scalar, vector and isovector, respectively.

Using the mean-field approximation to the Lagrangian density~in Eq. (\ref{Lag}) and the Legendre transformation, we obtain the mean-field Hamiltonian density. Applying the variational method and the Bogoliubov transformation which couples the mean-fields and pairing correlations, we arrive at the relativistic Hartree-Bogoliubov equation.~\cite{Kucharek:1991arbi},
\begin{gather}
     \begin{pmatrix}
          h_D - \lambda & \Delta \\ - \Delta^* & -h^*_D + \lambda
     \end{pmatrix}
     \begin{pmatrix}
         U_k \\ V_k
     \end{pmatrix}
     = E_k \,
     \begin{pmatrix}
         U_k \\ V_k
     \end{pmatrix}
\end{gather}
where $E_k$ denotes the energy  of a quasiparticle state $k$, $U_k$ and $V_k$ are the quasiparticle wave functions, and $\lambda$ represents the Fermi energy. The Dirac Hamiltonian $h_D$ is given by
\be
    h_D = \bm{\alpha} \cdot \bm{p} \, + \, \beta \left(M+S(\bm{r})\right) \, + \, V(\bm{r})
\ee
and the scalar $S(\bm{r})$ and vector $V(\bm{r})$ potentials  can be expressed as
\begin{subequations}
\bea
   S(\bm{r}) &=& \alpha_S \rho_S \,+\, \beta_S \rho^2_S \,+\,
               \gamma_S \rho^3_S \,+\, \delta_S \Delta\rho_S, \label{sPot} \\
   V(\bm{r}) &=& \alpha_V \rho_V \,+\, \gamma_V \rho^3_V \,+\, \delta_V \Delta \rho_V
             \,+\, e A_0 \,   \nonumber  \\   && +\, \alpha_{TV} \tau_3 \rho_{TV} \,+\, \delta_{TV} \tau_3 \Delta \rho_{TV}  . \label{vPot}
\eea
\end{subequations}
The local densities $\rho_S(\bm{r})$, $\rho_V(\bm{r})$ and $\rho_{TV}(\bm{r})$ take the following form in terms of the quasiparticle wave functions
\begin{subequations}
\bea
\rho_S(\bm{r}) = \sum_{k>0} \, \bar{V_k}(\bm{r}) V_k(\bm{r}), \\
\rho_V(\bm{r}) = \sum_{k>0} \, V^\dag_k (\bm{r}) V_k(\bm{r}), \\
\rho_{TV}(\bm{r}) = \sum_{k>0} \, V^\dag_k (\bm{r}) \tau_3  V_k(\bm{r}).
\eea
\end{subequations}
In principle we can derive the  pairing potential for particle-particle channel from the Lagrangian density~in Eq. (\ref{Lag}), but for simplicity adopt the following form
\bea \label{pairing_potential1}
\Delta_{k k^{'}}(\bm{r},\bm{r^{'}}) = - \sum_{\tilde{k}\tilde{k}^{'}} \,
                                 V^{pp}_{k k^{'},\tilde{k}\tilde{k}^{'}} (\bm{r},\bm{r^{'}})
                                 \kappa_{\tilde{k}\tilde{k}^{'}} (\bm{r},\bm{r^{'}}),
\eea
where $k$, $k^{'}$, $\tilde{k}$ and $\tilde{k}^{'}$ denote the quasiparticle states and the pairing tensor is defined by $\kappa=V^{\ast}U^T$~\cite{A_textbook_of_Peter_Ring}.  For $V^{pp}$  we use the density-dependent zero-range pairing interaction
\bea
V^{pp} (\bm{r},\bm{r^{'}}) = \frac{V_0}{2} \left(1 - P^{\sigma} \right)
                             \delta (\bm{r} - \bm{r^{'}} )
                             \left(1 - \frac{\rho(\bm{r})}{\rho_{sat}} \right),
\eea
where $\rho_{sat}$ is the nuclear saturation density. The total energy of a nucleus can be expressed as~\cite{Gambhir:1989mp, A_textbook_of_Jie_Meng}
\begin{eqnarray}
E_{\rm tot}&=& \sum_{k>0}(\lambda_{\tau}\! - \! E_k)v_k^2 - E_{\mathrm{pair}} +E_{\mathrm{c.m.}}
-\int \mathrm{d}^3\mathbf{r} \left[ \frac{1}{2}\alpha_S\rho^2_S \right. \nonumber \\
 &+& \frac{1}{2}\alpha_V\rho_V^2 + \frac{1}{2}\alpha_{TV}\rho_{TV}^2 + \frac{2}{3}\beta_S\rho_S^3
         +\frac{3}{4}\gamma_S\rho_S^4 +\frac{3}{4}\gamma_V\rho_V^4 \nonumber \\
 &+& \! \left. \frac{1}{2} ( \delta_S\rho_S\Delta\rho_S \!
         + \! \delta_V\rho_V\Delta\rho_V\! +\! \delta_{TV}\rho_3\Delta\rho_3\! +\! \rho_peA^0) \right]\, ,
\end{eqnarray}
where $E_{\mathrm{c.m.}}$ denotes the center of mass energy. The zero-range pairing force gives a local pairing field $\Delta(\mathbf{r})$ and the corresponding pairing energy is given by
\be
E_{\mathrm{pair}} = -\frac{1}{2}\int\mathrm{d}^3\mathbf{r} \kappa(\bm{r})\Delta(\bm{r}) .
\ee
To explore exotic nuclear properties, it is important to include self-consistently both the continuum and deformation effects and the coupling among them. We expand the wave functions in the Dirac Wood-Saxon (WS) basis~\cite{Zhou:2003jv} for the effects of continuum. To consider axial deformation with spatial reflection symmetry, we expand the  potentials ($S(\bm{r})$, $V(\bm{r})$) and densities ($\rho_S(\bm{r})$, $\rho_V(\bm{r})$, $\rho_{TV}(\bm{r})$) in terms of Legendre polynomials~\cite{Price:1987sf},
\be
f(\bm{r}) = \sum_{\lambda} \, f_{\lambda} (r) P_{\lambda} (\rm{cos}\theta),
            \,\,\, \lambda = 0, \, 2, \, 4, \, \cdots.
\ee
For the numerical calculation, we use the angular momentum cutoff for the Dirac WS basis $J_{\max}=23/2 \, \hbar$,
the expansion order $\lambda_{\max}=8$, the radius of the box $R_{\rm box}=20$ fm, and the energy cutoff for the Dirac WS basis $E_{\mathrm{cut}}=300$ MeV. For the particle-hole channel we adopt the  PC-PK1 density functional~\cite{Zhao:2010hi}, and for the particle-particle channel we use the zero-range pairing force with saturation density $\rho_{\rm sat}=0.152 \, \rm{fm^{-3}}$~\cite{Xia:2017zka} and the pairing strength $V_0 = -325 \,\, \rm{MeV}\,\rm{fm}^{\rm{3}}$. For more details of the numerical implementation of DRHBc theory, refer to~\cite{Zhang:2020wvp}. To obtain solid results, we first perform unconstrained calculations using different initial deformation parameter values, $\beta = -0.4, -0.3,\ldots, 0.5, 0.6$. 
In case we end up with more than one solution with very similar energies in unconstrained calculations, we check the potential energy curve made by the DRHBc calculation with constraints on the quadrupole deformation~\cite{Staszczak:2010zt}. This potential energy curve is used to study the shape coexistence.


\section{Bubble structure and shape coexistence} \label{chap_results}



\begin{figure*}[t]
\centering
\subfigure{\includegraphics[width=.4\textwidth]{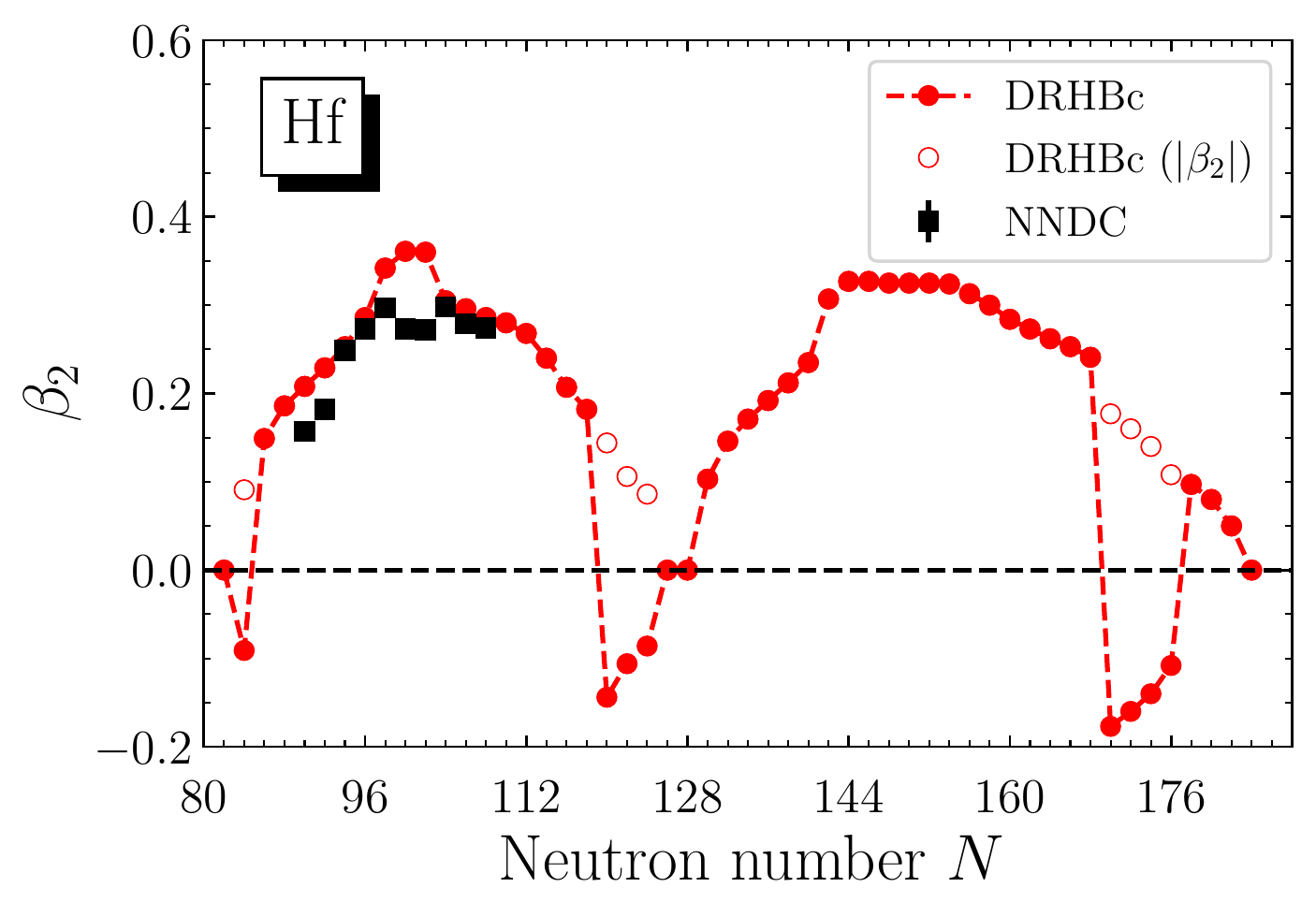} \label{Hf_isotopes_deformation}}
\subfigure{\includegraphics[width=.4\textwidth]{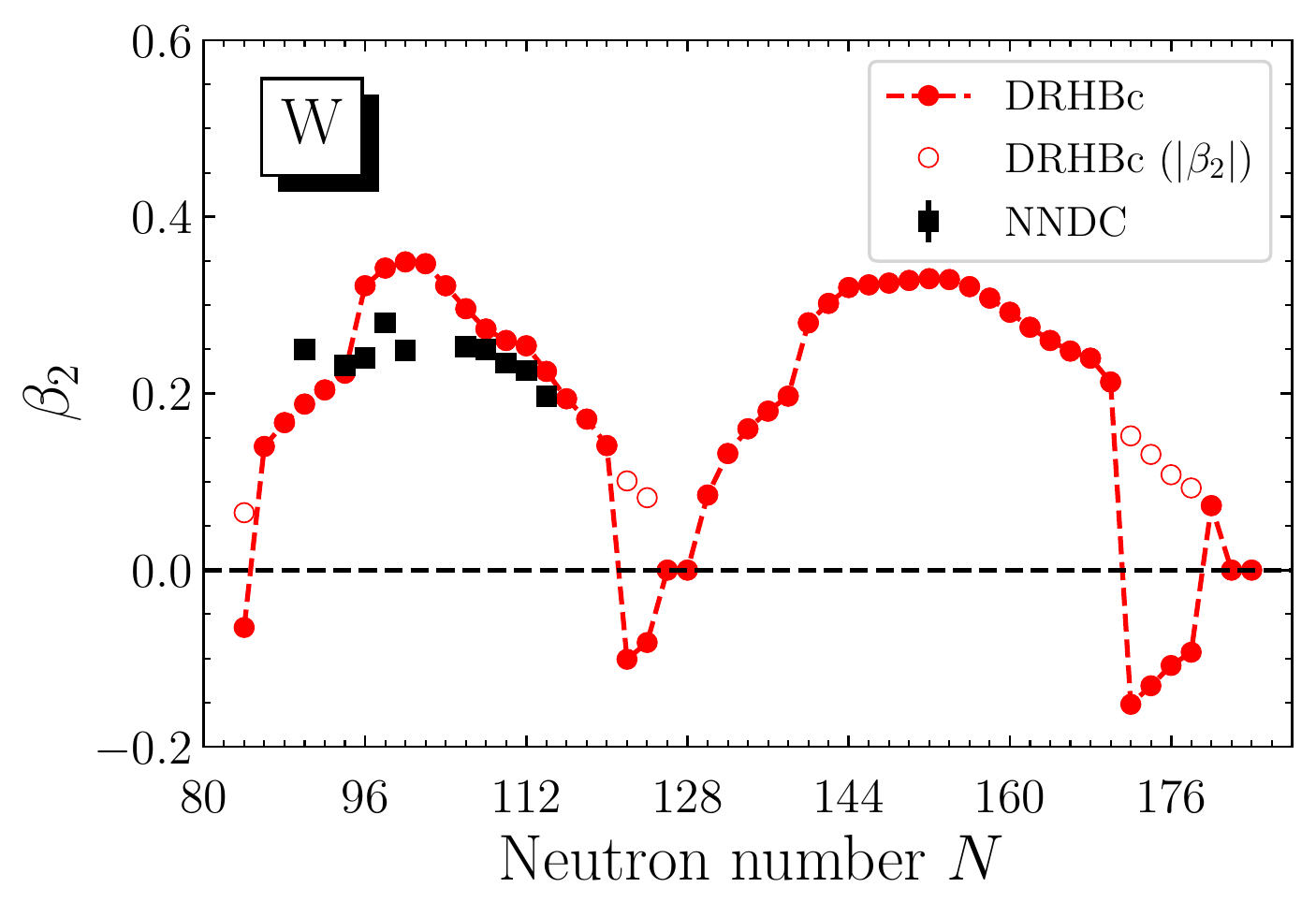} \label{W_isotopes_deformation}}
\subfigure{\includegraphics[width=.4\textwidth]{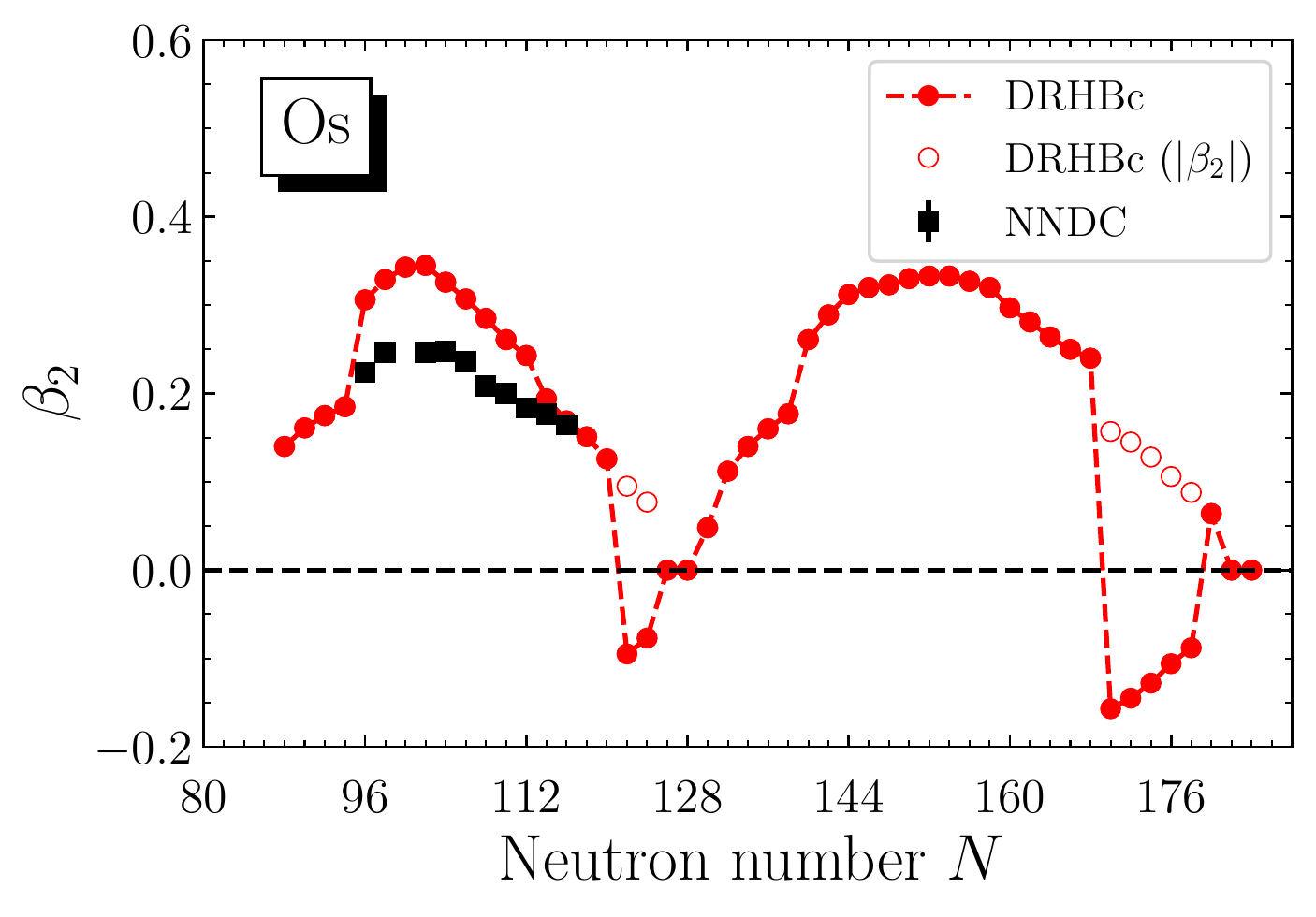}\label{Os_isotopes_deformation}}
\subfigure{\includegraphics[width=.4\textwidth]{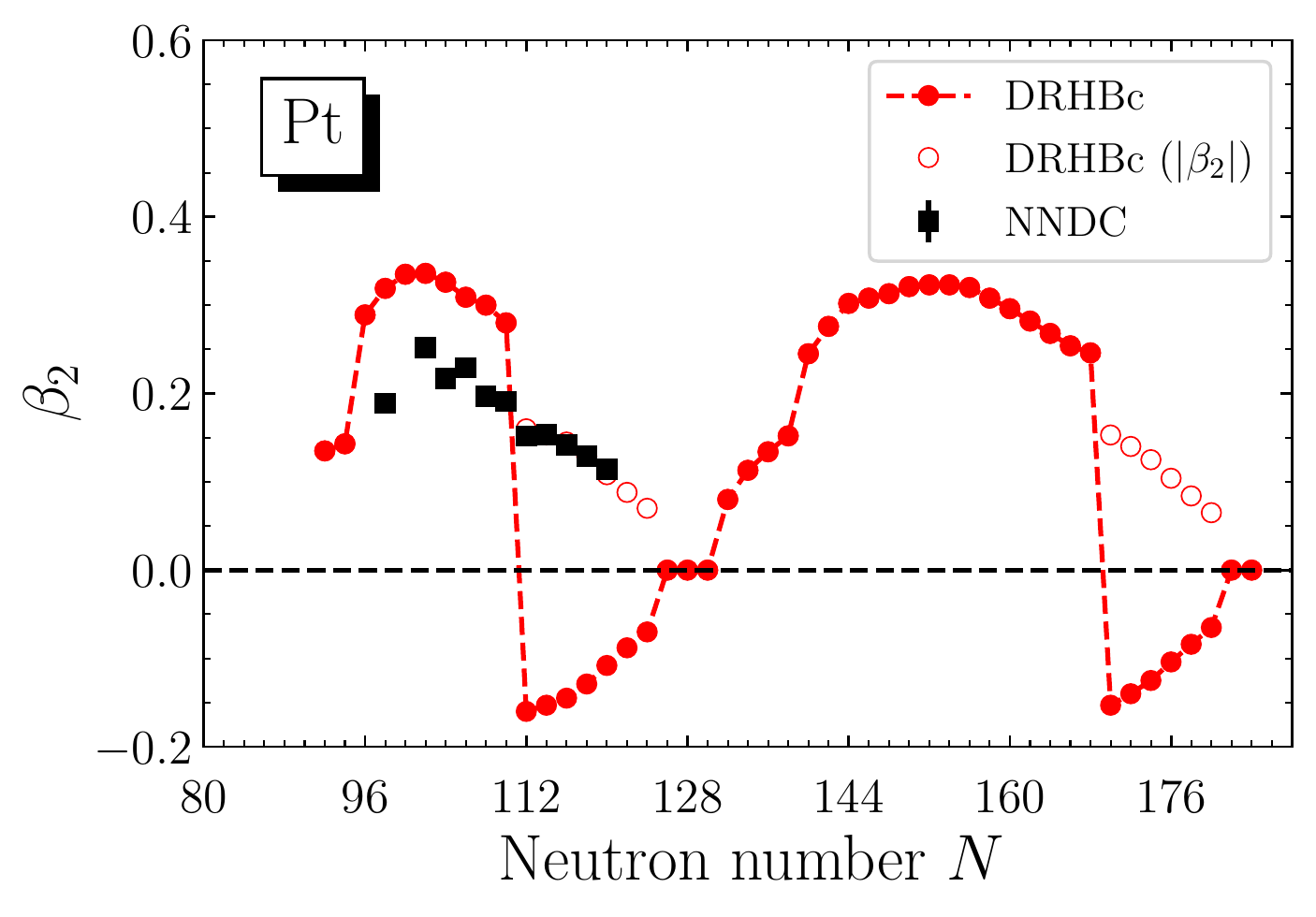}\label{Pt_isotopes_deformation}}
\subfigure{\includegraphics[width=.4\textwidth]{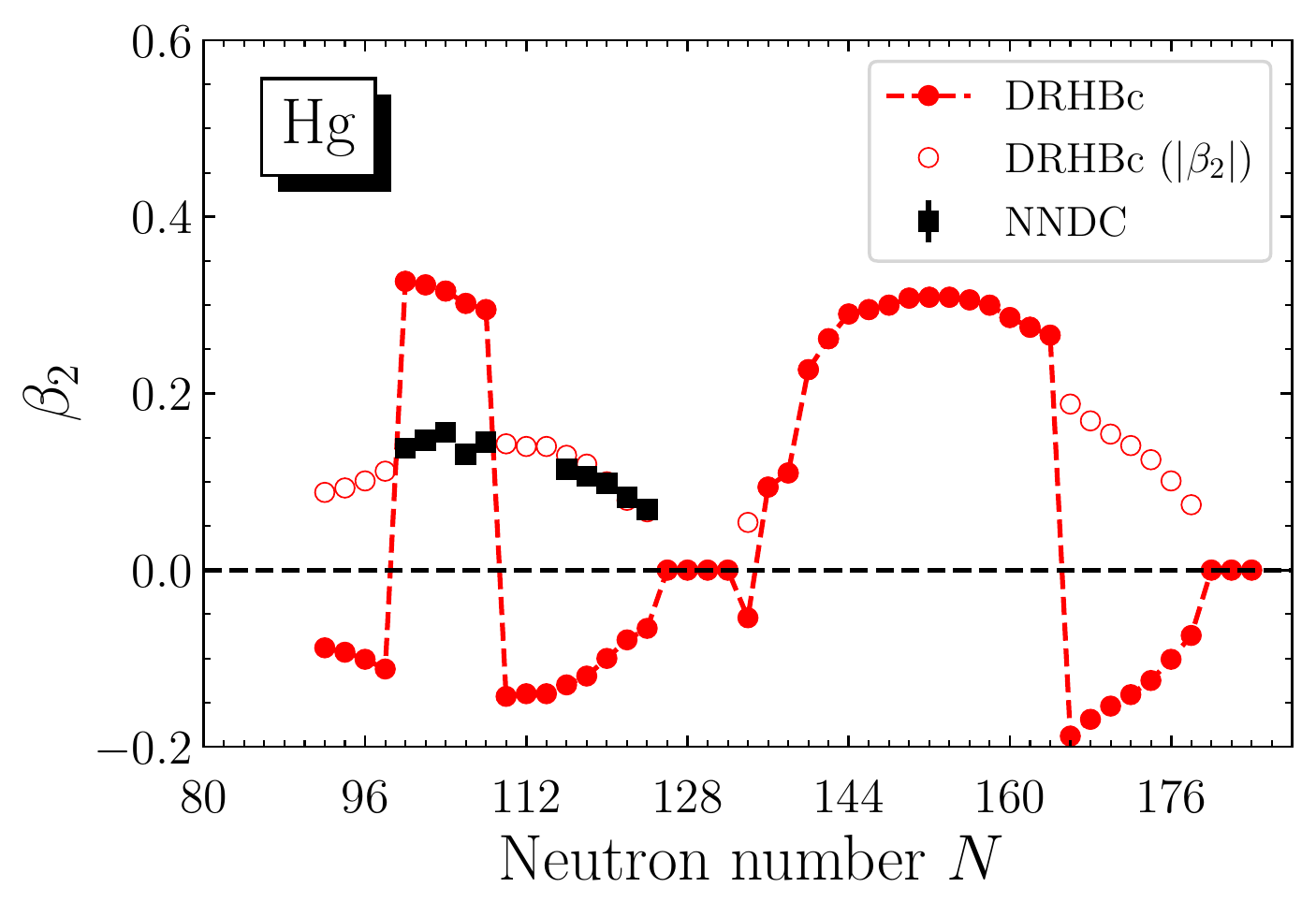}\label{Hg_isotopes_deformation}}
\caption{ The quadrupole deformations of Hf, W, Os, Pt and Hg isotopes as function of neutron number. }
\label{isotopes_deformation}
\end{figure*}

Before presenting  our results on nuclear bubble structure and shape coexistence, we first show the quadrupole deformations of the isotopes considered in this work and compare them with experiments since the deformation and pairing affect a lot the formation of bubble nuclei~\cite{Khan:2007ji, Luo:2018waj, Saxena:2019ish}. The quadrupole deformation $\beta_2$ can be obtained
by calculating the quadrupole moment $Q_2$ and root-mean-square radius $\langle r^2 \rangle$ as follows,
\bea
\beta_2 &=& \frac{\sqrt{5\pi} Q_2}{3 N \langle r^2 \rangle}, \label{beta2} \\
Q_2 &=& \frac{16\pi}{5} \langle r^2 Y_{20} (\theta, \phi) \rangle, \label{Q2} \\
\langle r^2 \rangle &\equiv& \frac{\int d^3 \mathbf{r} ~ \bm{r}^2 \rho(\bm{r}) }{N}, \label{rms_mean}
\eea
where $N$, $Y_{20}, \rho(\vec{r})$ are the number of nucleon, spherical harmonic function of degree 2  and order 0, and nucleon density, respectively.
Fig. \ref{isotopes_deformation}  shows the quadrupole deformation for Hf, W, Os, Pt, and Hg isotopes. Here, experimental and evaluated data are taken from NNDC \cite{NNDC}. As it should be, the quadrupole deformation values of the nucleus at the neutron magic number ($N=82,~ 126, ~184$) are zero. Also, the quadrupole deformations predicted in the present work are in qualitative agreement with those from NNDC.

Finally, we briefly discuss if the neutron drip line changes due to deformation effects.
Since the last-bound isotopes in DRHBc calculations are magic nuclei at $N=184$,
the neutron drip lines of Hf, W, Os, Pt, and Hg isotopes in RCHB and in DRHBc are the same.
For an extensive summary of nuclear properties including the isotopes discussed in this work, we refer to
\cite{KZhangetal}.

\subsection{Bubble structure \label{bubble-C} }

As well known,  the central-depleted density of a spherical nucleus is due to the unoccupancy of the $l$=0 state (${\rm s}_{1/2}$ orbitals) near the Fermi surface. In the case of deformed nuclei, due to the mixing of $s$ orbitals with higher angular momentum orbitals bubble structure can be hindered~\cite{Shukla:2014bsa, Yao:2012cx, Luo:2018waj}. It was also observed that not only deformations but also pairing correlations disfavor bubble structures~\cite{Khan:2007ji, Grasso:2009zza, Saxena:2019ish, Yao:2012cx, Wang:2015coa}.

To investigate the bubble nuclei, as in a previous work~\cite{Saxena:2019ish}, one can define the bubble parameter which corresponds to the proton depletion fraction
\be
{\cal B}_p  \equiv \bigg( 1- \frac{\rho_{p, \rm c}}{\rho_{p, \rm max}} \bigg) \times100~[\%] \label{depletion_fraction_definition}.
\ee
where $\rho_{p,\rm c}$ is the central proton density and $\rho_{p,\rm max}$ is the maximum proton density in nuclei. There is no clear-cut criterion for defining a bubble nucleus. In this work we set the criterion for determining the bubble nucleus as ${\cal B}_p = 20\%$. Note that, in the previous work, ${\cal B}_p \simeq 15\%$ is used to find the bubble nuclei \cite{Saxena:2018qgd}. As described in Sec.~\ref{chap_DRHBc_framework}, the nucleon density $\rho(\bm{r})$ is expanded in terms of Legendre polynomials~\cite{Price:1987sf},
\be
\rho(\bm{r}) = \sum_{\lambda} \, \rho_{\lambda} (r) P_{\lambda} (\rm{cos}\theta),
            \,\,\, \lambda = 0, \, 2, \, 4, \, \cdots.
\ee
For nearly spherical nuclei, the spherical symmetric component of the density $\rho_{\lambda=0}(r)$ dominates the density over higher orders in $\lambda$ and one can easily obtain the maximum density from $\rho_{\lambda=0}(r)$. Consequently, the spherically symmetric maximum density shell can be easily obtained. In previous works~\cite{Friedrich:1982esq,Schuetrumpf:2017qeq,Saxena:2019ish}, instead of $\rho_{p, \rm max}$, the average proton density of the nucleus assuming a constant density up to the diffraction radius is also used to study bubble structures of spherical nuclei. However, since highly deformed nuclei are included in our work, we did not consider the bubble parameter with diffraction radius.

\begin{table}[b]
\begin{center}
\caption{Candidates of bubble nuclei with top 10 highest bubble paramters. The listed nuclei are spherical except for
$^{ 254 }$Hf which is slightly deformed.
} \label{hiDF-1}
\begin{tabular}{|c|c|c|c|}
\hline
  \multirow{2}{*}{ Nucleus }           & \multicolumn{2}{c|}{DRHBc}  & RCHB \\
     \cline{2-4}
          & \phantom{xx} $\beta_2$ \phantom{xx}       & \phantom{xx} ${\cal B}_p^\star$ [$\%$] \phantom{xx}
          &  ${\cal B}_p^\star ( = {\cal B}_p)$  [$\%$]\\
\hline
$^{256}$Hf & 0.000 & 29.2   & 27.4 \\
\hline
$^{258}$W & 0.000 & 28.3    & 26.6  \\
\hline
$^{260}$Os & 0.000 & 27.2  & 25.5 \\
\hline
$^{256}$W & 0.000 & 26.7  & 25.1 \\
\hline
$^{258}$Os & 0.000 & 26.0   & 24.3  \\
\hline
$^{254}$Hf & 0.057 & 25.7   & 25.7  \\
\hline
$^{200}$Hf & 0.000 & 25.6   & 24.8 \\
\hline
$^{198}$Hf & 0.000 & 25.3   & 24.5  \\
\hline
$^{262}$Pt & 0.000  & 25.2  & 23.5  \\
\hline
$^{202}$W & 0.000 & 25.1   & 24.3  \\
\hline
\end{tabular}
\end{center}
\end{table}

\begin{table*}[b]
\begin{center}
 \caption{List of some deformed bubble candidates. Densities $\bar \rho_{p,\rm max}$ and $\rho_{p, \rm max}$ are in fm$^{-3}$.
  \label{deformedB-1} }
  \begin{tabular}{|c|c|c|c|c||c|c|c|c|}
    \hline
     \multirow{2}{*}{ Nucleus }             & \multicolumn{6}{c|}{DRHBc}  & DRHBc$^\dagger$ & RCHB \\
     \cline{2-9}
     & $\beta_2$ & $\rho_{\rm c}$ & $\bar{\rho}_{p,\rm max}$ & \phantom{xx} ${\cal B}_p^\star$ [$\%$] \phantom{xx}
     & $\rho_{p,\rm max}$ & \phantom{xx} ${\cal B}_p$ [$\%$] \phantom{xx}  &   $ {\cal B}^\star_{p} |_{\beta_2=0}$ [$\%$]
     &   ${\cal B}_p^\star (={\cal B}_p)$ [$\%$] \\
    \hline
    $^{254}$Hf & 0.057 & 0.0350 & 0.0471 & 25.7 & 0.0482 & 27.4 & 27.3 & 25.7  \\
    \hline
    $^{228}$Pt & 0.321 & 0.0417 & 0.0556 & 25.1 & 0.0578 & 27.8 & 30.1 & 29.0 \\
    \hline
    $^{230}$Hg & 0.308 & 0.0422 & 0.0562 & 24.9 & 0.0582 & 27.5 & 27.0 & 26.0\\
    \hline
    $^{252}$Hf & 0.074 & 0.0359 & 0.0475 & 24.4  & 0.0492 & 27.1 & 25.8 & 24.2 \\
    \hline
    $^{254}$W & 0.068 & 0.0365 & 0.0481 & 24.2 & 0.0497 & 26.6 & 25.2 & 23.6 \\
    \hline
    $^{206}$Os & 0.048 & 0.0451 & 0.0593 & 24.1 & 0.0600 & 24.9 & 24.8 & 24.0 \\
    \hline
    $^{204}$W & 0.080 & 0.0447 & 0.0586 & 23.8 & 0.0600 & 25.6 & 25.4 & 24.6 \\
    \hline
    $^{202}$Hf & 0.100 & 0.0442 & 0.0580 & 23.7 & 0.0601 & 26.4 & 25.9 & 25.1 \\
    \hline
  \end{tabular}\\
   $^\dagger$ Results of constrained DRHBc calculation ($\beta_2=0$ fixed).
     \end{center}
\end{table*}

For deformed nuclei,  one has to include higher order terms in $\lambda$. The maximum proton density for a given azimuthal angle $\theta$, $\rho_{p,\rm max} (r ; \theta)$, is not a constant and the maximum proton density shell is deformed. In this work, in order to implement the effects of deformations, we define an averaged maximum proton density
\be
\bar{\rho}_{p,\rm max} = \frac{\int \rho_p (r, \theta) \delta(r- r_{\rm max}(\theta)) dV}
{ \int   \delta(r- r_{\rm max}(\theta) ) dV}\label{DFD}\, .
\ee
where $r_{\rm max} (\theta)$ is the radius of the maximum density for a given angle $\theta$. In practice, the average can be taken over the deformed 2-dim maximum density shell because of the delta function. With $\bar\rho_{p,\rm max}$ we define a modified bubble parameter
 \be
{\cal B}_p^\star  \equiv \bigg( 1- \frac{\rho_{p, \rm c}}{\bar \rho_{p, \rm max}} \bigg) \times100\% \label{modified_bubble_parameter}.
\ee
Note that $\bar \rho_{p,\rm max} = \rho_{p,\rm max}$ and $B_p^\star = B_p$ for a spherical nucleus.

In Table~\ref{hiDF-1}, candidates of bubble nuclei with the top 10 highest bubble parameters in this study are summarized. In the table, we also compare bubble parameters from the spherical relativistic continuum Hartree–Bogoliubov (RCHB) theory~\cite{Xia:2017zka}. From Table~\ref{hiDF-1} we can infer that deformation may weaken bubble structures because almost all the nuclei with high bubble parameters are spherical except for $^{254}$Hf which is slightly deformed. Also, by comparing our spherical ($\beta_2=0$) results with those from the RCHB, one can observe the role of the pairing in the formation of bubble structure since one of the main differences in the parameters between the DRHBc and the RCHB is the paring strength.
The pairing strength in the DRHBc is $-325.0$ MeV ${\rm fm}^3$ and that in the RCHB is $-342.5$ MeV ${\rm fm}^3$, which implies that the pairing effect is stronger in the RCHB. For the comparison of the pairing energies between the DRHBc calculations and RCHB calculations, see Fig. 13 in Ref.~\cite{Xia:2017zka}. Since the bubble parameters in RCHB (wih bigger pairing strength) is smaller than those in DRHBc, we may conclude that the pairing effects hinder the formation of bubble nuclei.
Note, however, that there are some other differences between RCHB and DRHBc;
the angular momentum cutoffs are $(19/2)\hbar$ in RCHB and $(23/2)\hbar$ in DRHBc. In addition, the method to solve the Dirac equations are different in RCHB and DRHBc. In DRHBc, Dirac equations is solved by expanding in Dirac WS basis, but in RCHB, Dirac equations is solved using shooting method in the coordinate space. Therefore, even if the same pairing strength is used in both theories, the effect of the pairing strength may not be the same
since the pairing effect is related to the truncation of the WS basis in the DRHBc. The effect of the pairing strength can be checked by artificially increasing the pairing strength in the DRHBc calculation. For example, in the case of $^{256}$Hf, if the pairing strength -342.5 MeV fm$^3$ is used in the DRHBc calculation, the modified bubble parameter ${\cal B}_p^\star=29.2\%$ is shifted to 27.2$\%$. This is consistent with the results in other works~\cite{Khan:2007ji, Grasso:2009zza, Saxena:2019ish, Yao:2012cx, Wang:2015coa}.

\begin{figure*}[t]
\centering
\subfigure{\includegraphics[width=.4\textwidth]{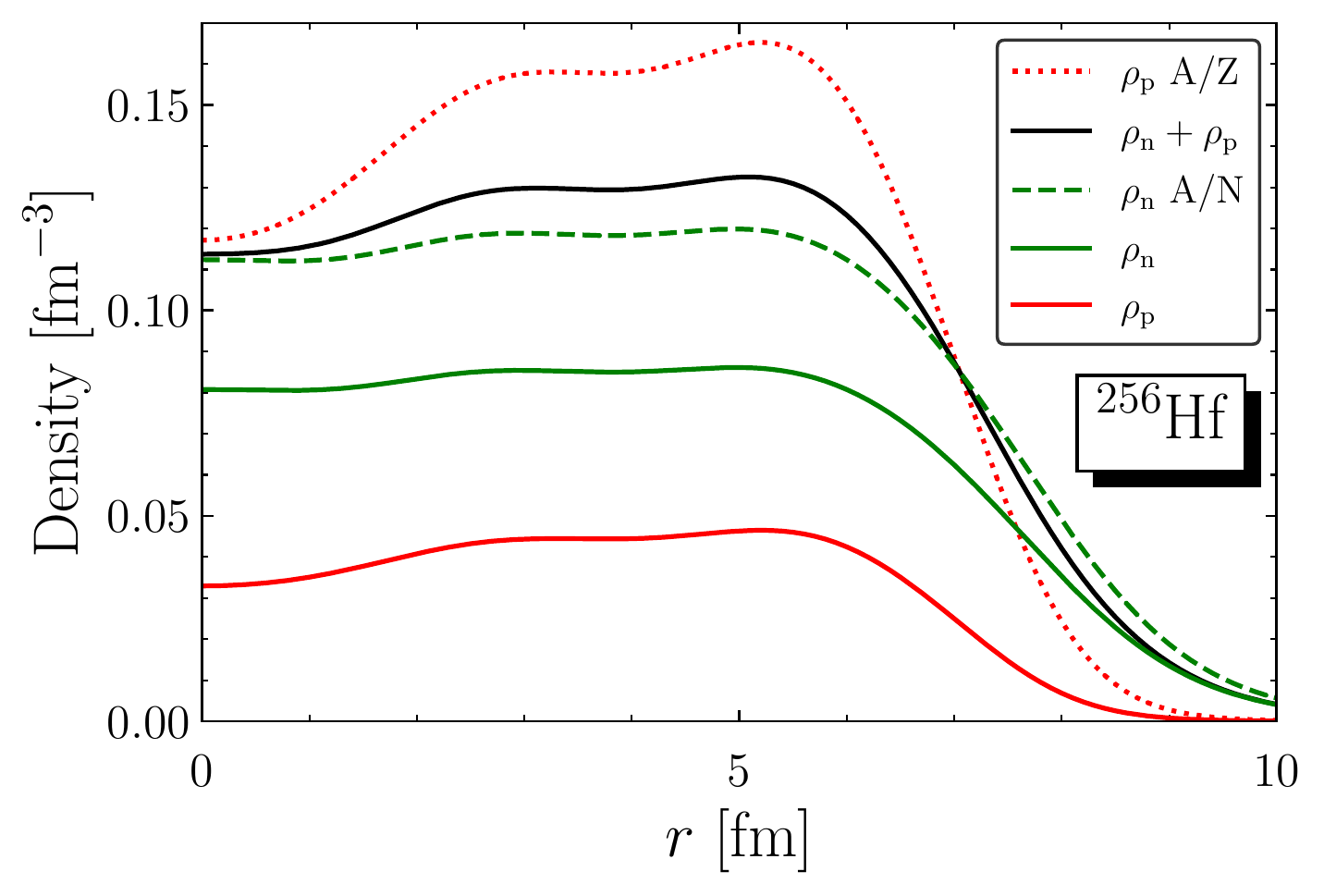}}
\subfigure{\includegraphics[width=.4\textwidth]{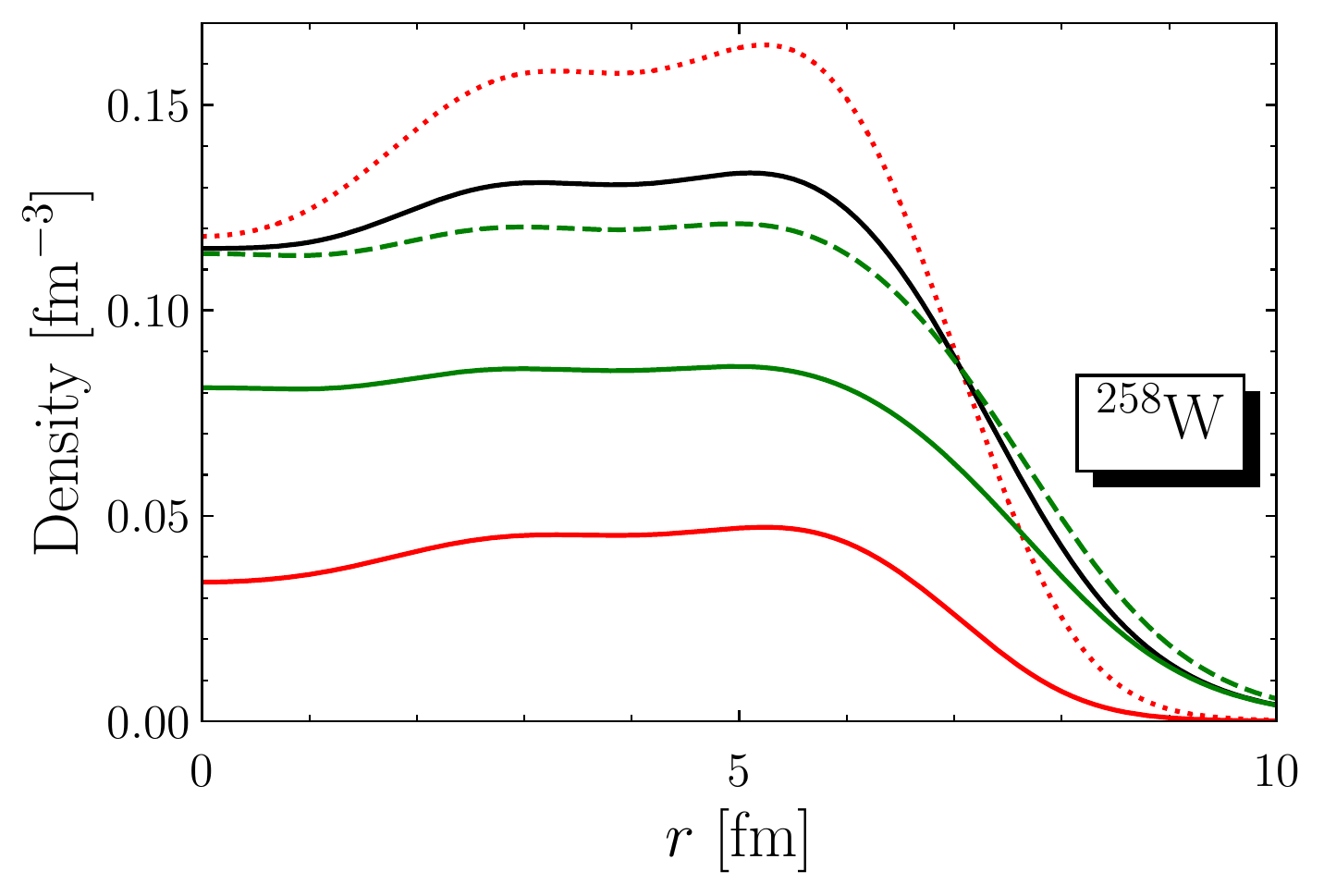}}
\subfigure{\includegraphics[width=.4\textwidth]{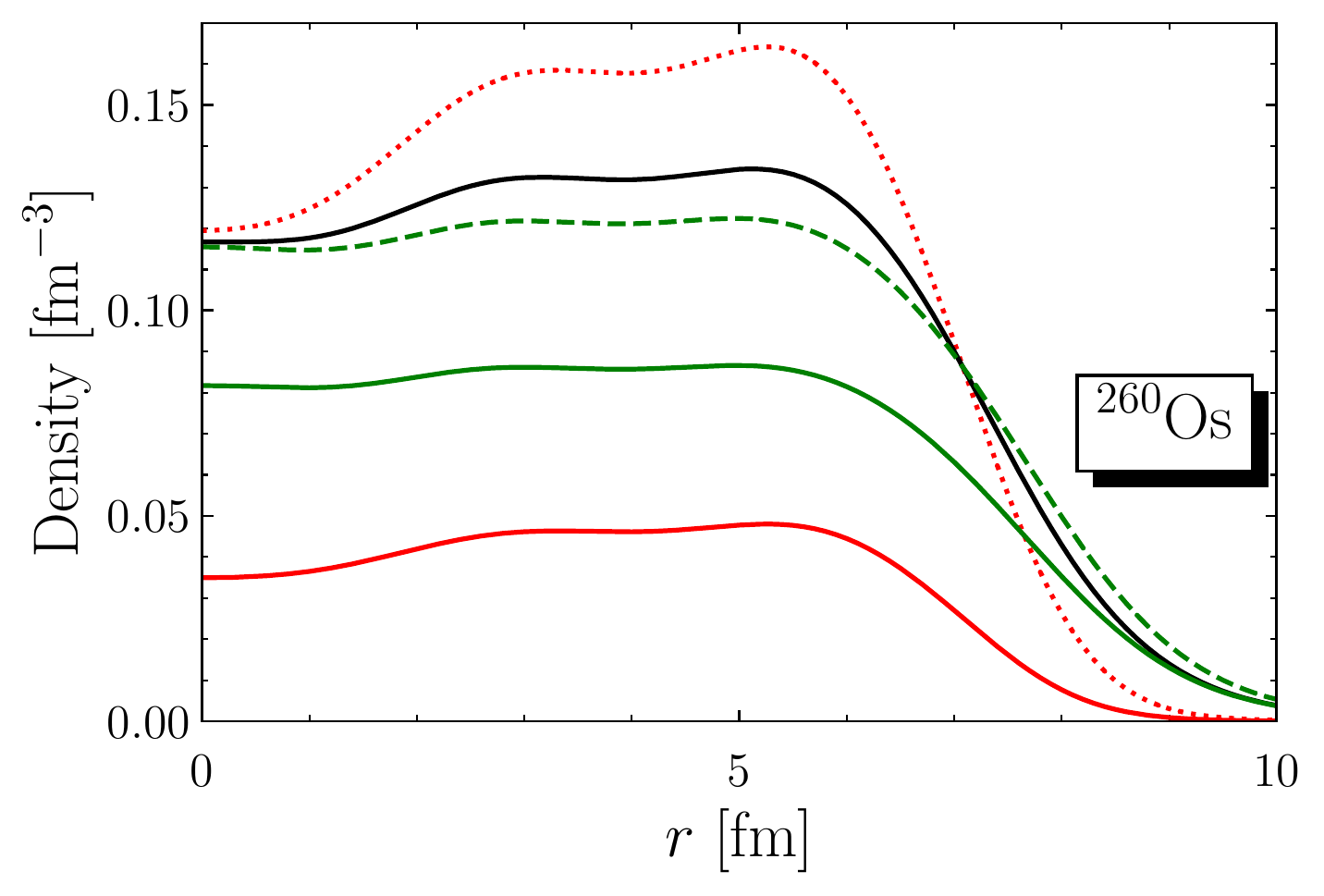}}
\subfigure{\includegraphics[width=.4\textwidth]{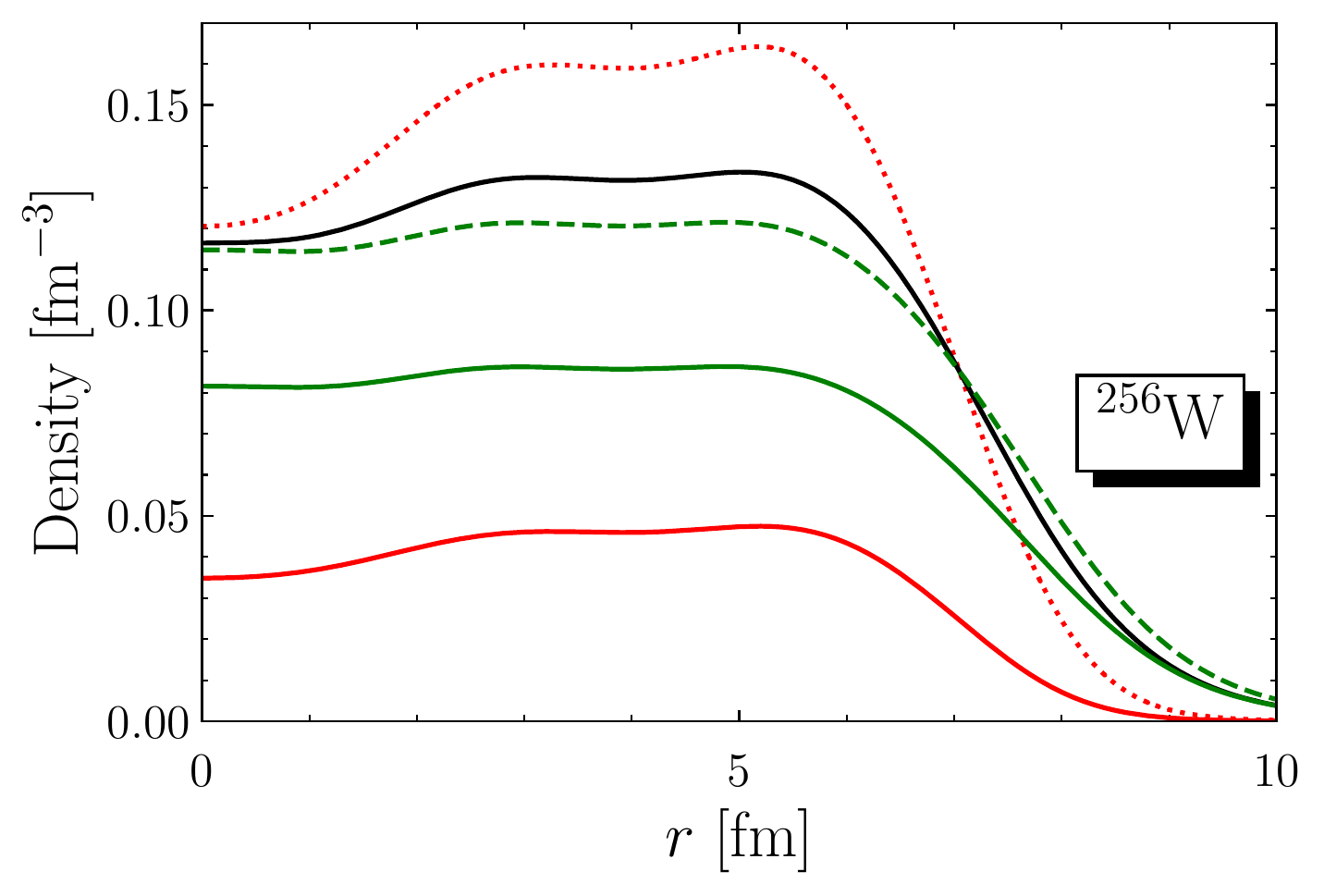}}
\caption{(Scaled) neutron density, (scaled) proton density, and total nuclear density profiles of most central-depleted nuclei. }
\label{density_profiles}
\end{figure*}

\begin{figure*}[t]
\centering
\subfigure{\includegraphics[width=.4\textwidth]{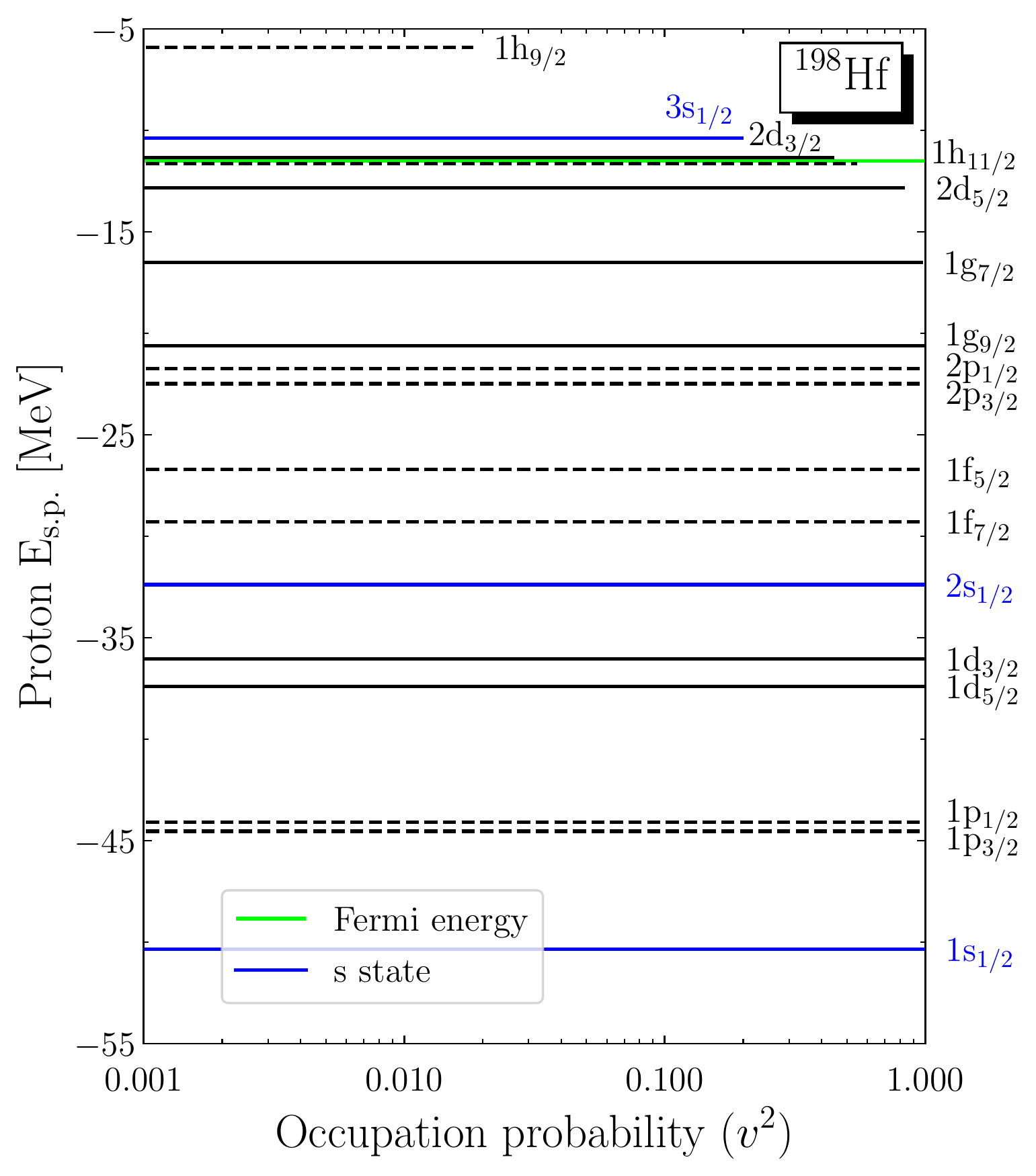}}
\subfigure{\includegraphics[width=.4\textwidth]{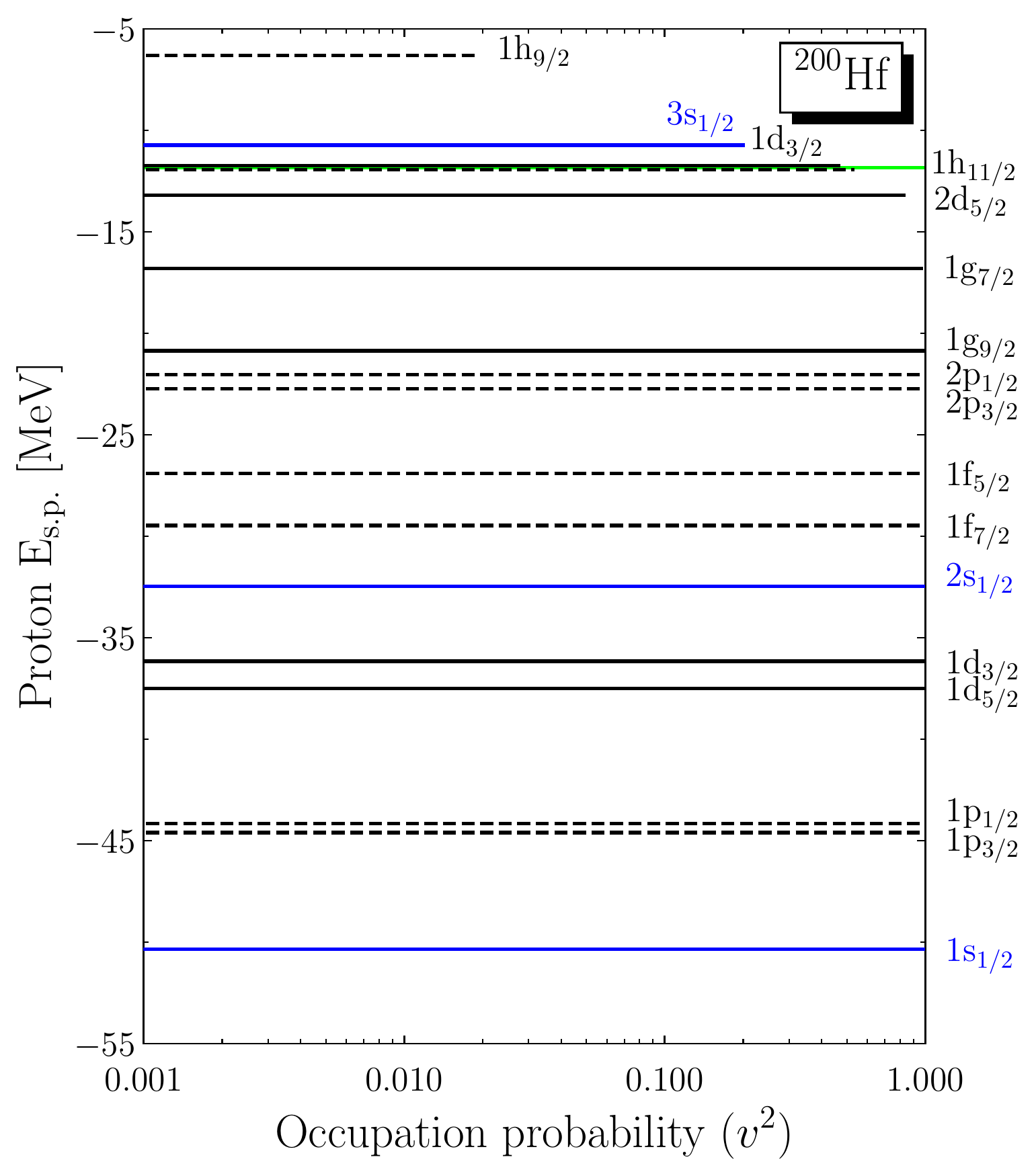}}
\subfigure{\includegraphics[width=.4\textwidth]{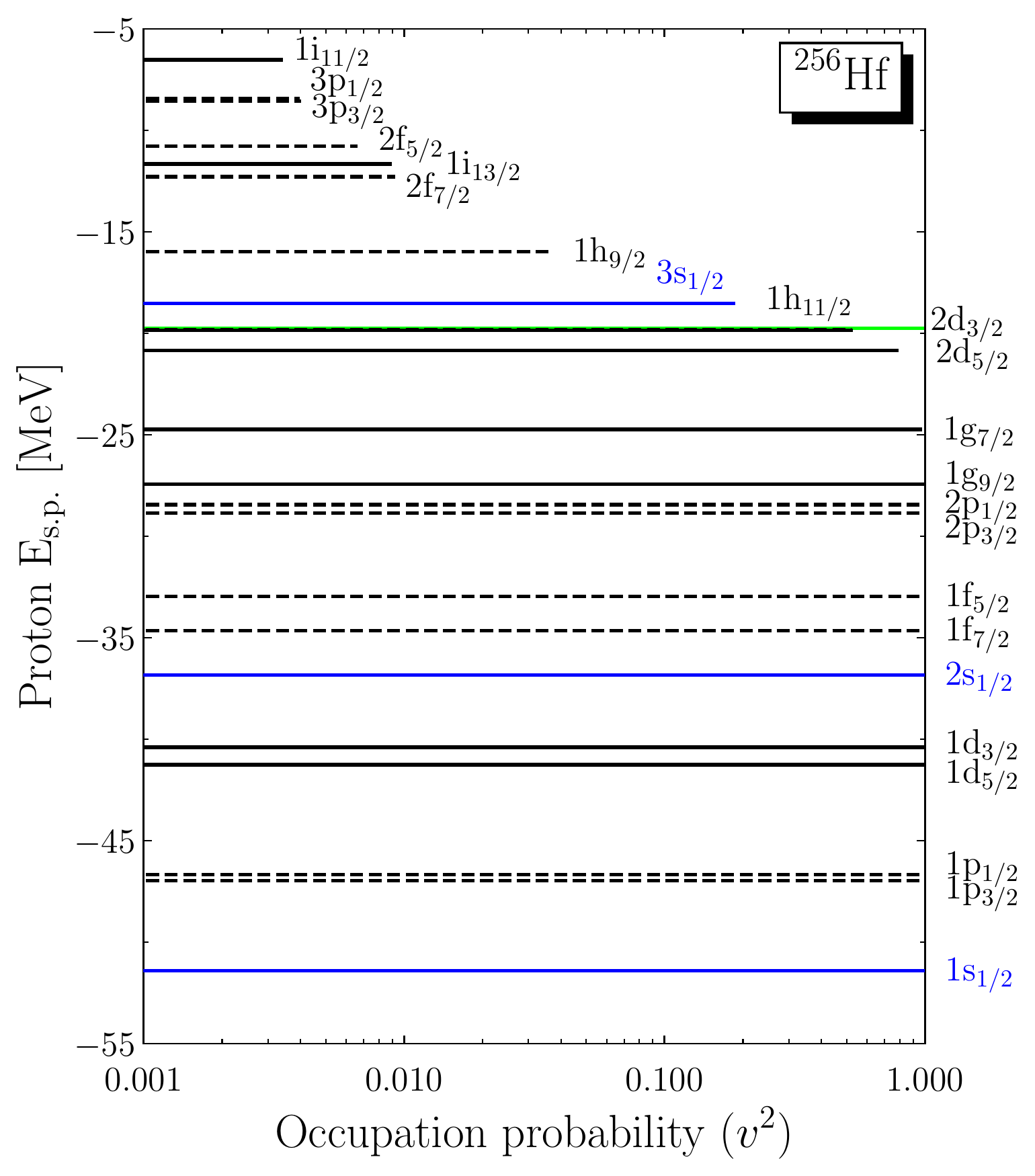}}
\caption{Proton single-particle levels around the Fermi surface (green dashed line) for $^{198}$Hf, $^{200}$Hf and $^{256}$Hf. Blue lines represent $l=0$ state and the grey lines represent the other states. The solid lines represent positive parity states and the dashed lines denote the negative parity states. The occupation probability of the s-states near the Fermi surface is low, which reduces the central density and induces bubble structure. }
\label{Hf_ground_energies}
\end{figure*}

Figure \ref{density_profiles} shows the density profiles of the most central depleted nuclei.
To illustrate the bubble structure of the proton density, we plot the scaled neutron ($\rho_\text{n}$A/N) and scaled proton density ($\rho_\text{n}$A/Z) in order to guide the eye, compared with the total baryon density.
We plot the proton single-particle energy levels of spherical Hf isotopes with corresponding single-particle occupation probabilities  in Fig.~\ref{Hf_ground_energies}. For spherical nuclei, the angular momentum $l$ is a good quantum number, and only $l$=0 states contribute to the central density. The low occupation probability of s orbitals near the Fermi energy leads proton central density depleted. For deformed nuclei, the states in spherical nuclei are splitted (see Fig~\ref{Os_Nilsson_diagrams}). $1/2^+$ states from the splitted orbitals could contribute to central density and bubble structure could be hindered in deformed nuclei, for example, see~\cite{Khan:2007ji, Grasso:2009zza, Saxena:2019ish, Yao:2012cx, Wang:2015coa}.

In Table~ \ref{deformedB-1}, the candidates of deformed bubble nuclei are summarized. In the table, in order to investigate the effects of the deformation closely, we compare our results of deformed bubble candidates with those of the RCHB. If we consider ${\cal B}_p^\star$, the deformation tends to lower the bubble parameters of the isotopes, especially for the isotopes with large $\beta_2$ values such as $^{228}$Pt. In order to see the density fluctuations within the maximum density shell, described near Eq.~(\ref{DFD}), we compare  $\bar\rho_{p, \rm max}$ (${\cal B}_p^\star$) and  $\rho_{p,\rm max}$ (${\cal B}_p$) in Table~ \ref{deformedB-1}. For the deformed bubble candidates, the density fluctuations within the deformed shell are non-negligible and cause ${\cal O}(10\%)$ differences in the depletion fraction.

Since the pairing strengths in the DRHBc and RCHB are different, we discuss the competition between pairing and deformation effects. While ${\cal B}_p^\star$ of DRHBc are larger than those of RCHB in Table~\ref{hiDF-1} (spherical cases), the results are opposite in Table~\ref{deformedB-1} (deformed cases). We notice here that for the isotopes with larger $\beta_2$ values, the reduction in bubble parameters are bigger. For the comparison, we also add results of constrained DRHBc calculation ($\beta_2=0$) in Table~ \ref{deformedB-1}.  Since the bubble parameters in RCHB (wih bigger pairing strength) is smaller than those in DRHBc$^\dagger$, we confirm again that the pairing effects hinder  the formation of bubble nuclei as in Table~\ref{hiDF-1}.

\subsection{Shape coexistence}\label{scoex}

Shape coexistence is another important and interesting feature of  nuclei; nuclear shapes coexist within the tiny energy range of nuclear excitations. The almost degenerate minima are related to the low single-particle energy level density around the Fermi levels of the neutron or proton~\cite{Maharana:1992zz, Lalazissis:1998ew, Xiang:2011fb, Kumar:2020nza}. Yang et al.~\cite{Yang:2021bor} found that the degenerate minima are more related to quadrupole deformation than triaxial deformation.
Many experiments have been carried out to discover shape coexistence in nuclei, for example, see ~\cite{Bree:2014mxa, Rapisarda:2017hzb, Muller-Gatermann:2019ofd, Fortune:2019efj, Olaizola:2019hzn,Siciliano:2020akh} for Hg isotopes.

\begin{figure}[t]
\centering
\subfigure{\includegraphics[width=.4\textwidth]{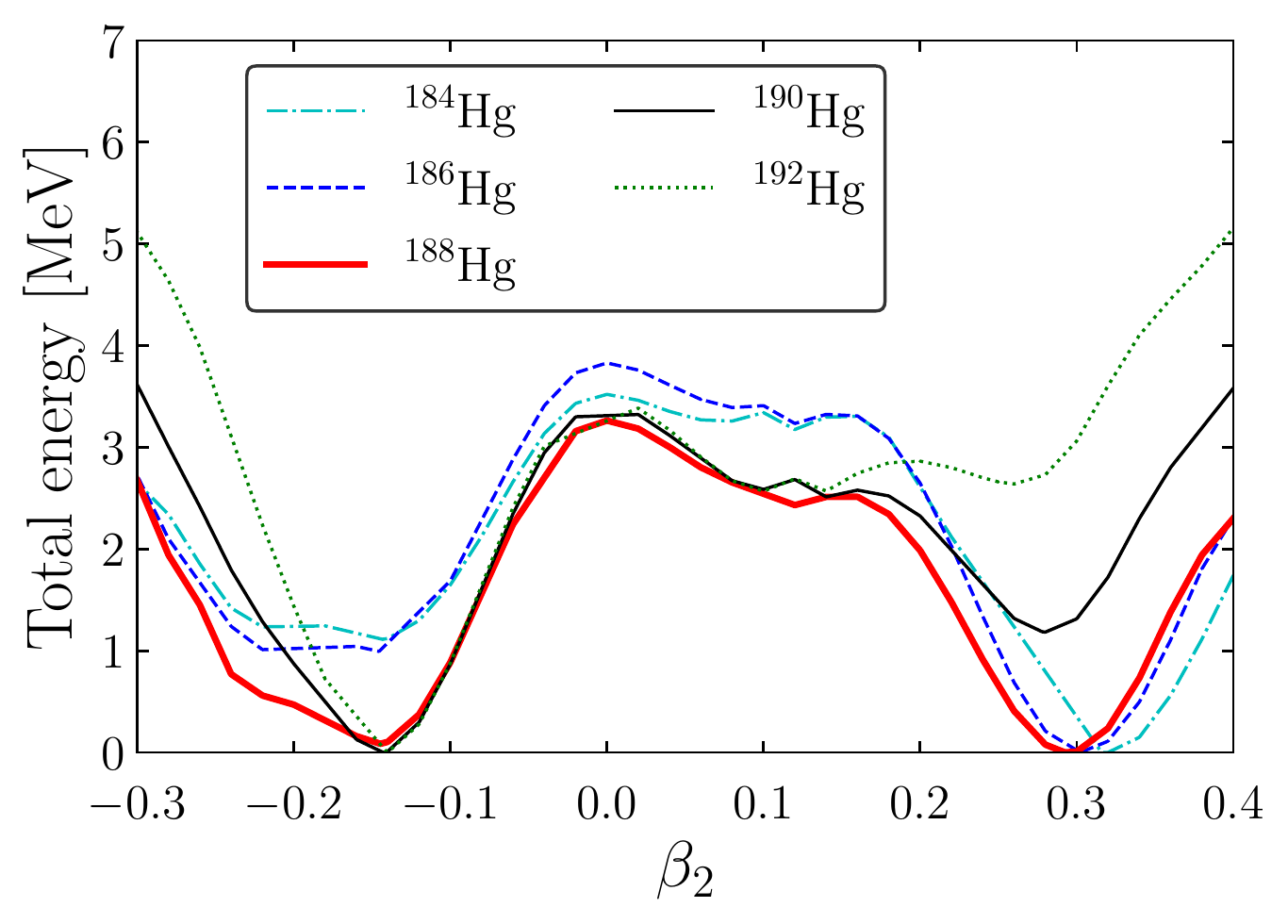}}
\subfigure{\includegraphics[width=.4\textwidth]{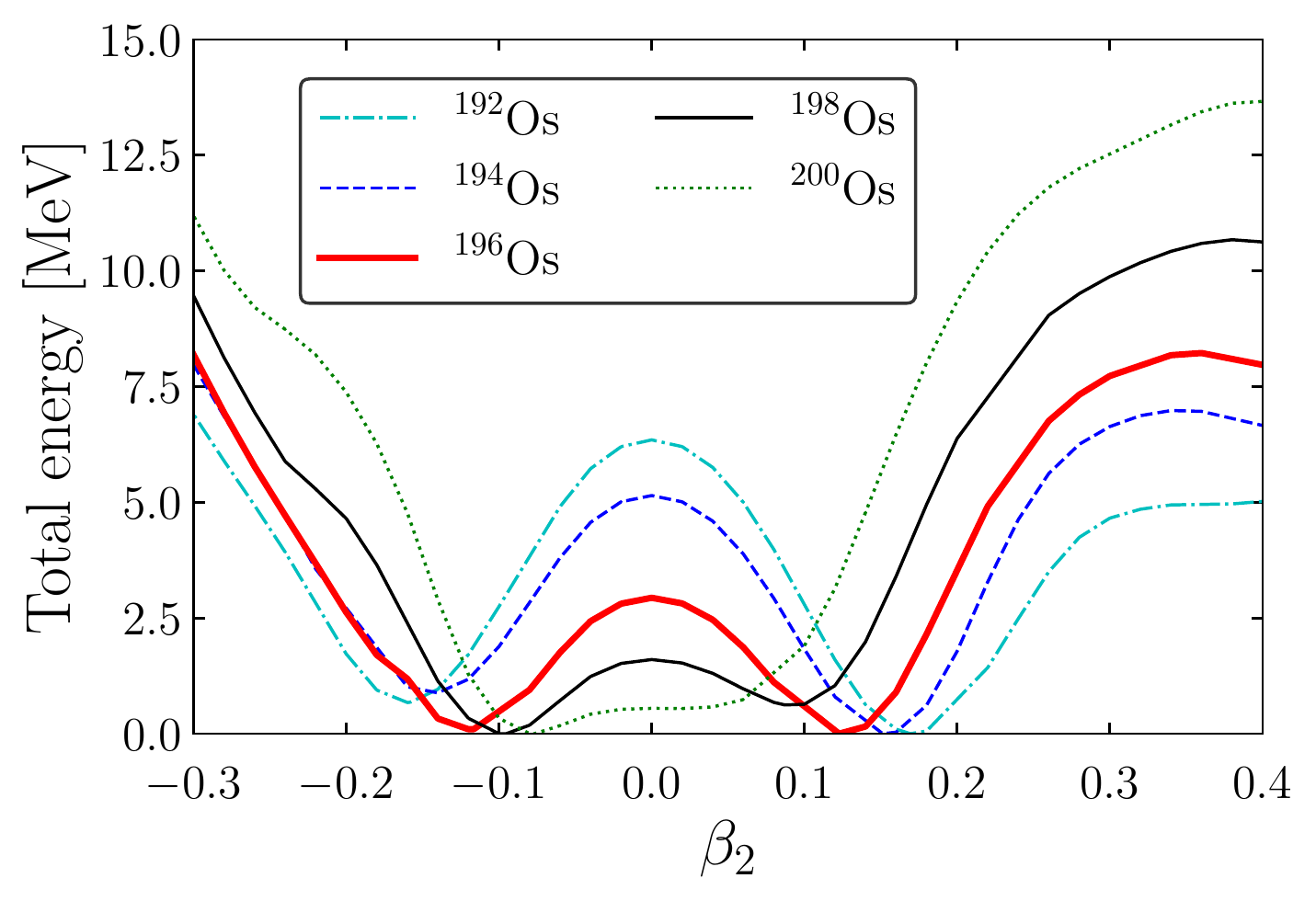}}
\caption{The potential energy curve of $^{184,186,188,190,192}$Hg and $^{192,194,196,198,200}$Os as a function of the quadrupole deformation $\beta_2$.}
\label{shape_coexistence_Hg_and_Os}
\end{figure}

To identify a nucleus with shape coexistence we first evaluate the energy of the different shape configurations in the framework of DRHBc and check the energy difference among them. To obtain precisely the almost degenerate minima, we perform both unconstrained calculations and constrained calculations, especially near the local minima. We then select candidate nuclei that exhibit different shapes with very small energy differences, roughly $|\Delta E|\lesssim 1$ MeV. As an example we plot the potential energy curves of  $^{184,186,188,190,192}$Hg and $^{192,194,196,198,200}$Os isotopes in Fig.~\ref{shape_coexistence_Hg_and_Os}. Among them $^{188}$Hg and $^{196}$Os are the candidates.

We find several candidates for each isotope chain even with a more strict criterion for $|\Delta E| \leq 0.5$  MeV: $^{156,}\,\!$ $^{190,}\,\!$ $^{192,}\,\!$ $^{202,}\,\!$ $^{234,}\,\!$ $^{236,}\,\!$ $^{240,}\,\!$ $^{242,}\,\!$ $^{248,}\,\!$ $^{250}$Hf, $^{158,}\,\!$ $^{194,}\,\!$ $^{204,}\,\!$ $^{244,}\,\!$ $^{252,}\,\!$ $^{254}$W, $^{196,}\,\!$ $^{206,}\,\!$ $^{208,}\,\!$ $^{246,}\,\!$ $^{254,}\,\!$ $^{256}$Os, $^{190,}\,\!$ $^{196,}\,\!$ $^{210,}\,\!$ $^{212,}\,\!$ $^{246,}\,\!$ $^{258}$Pt and $^{172,}\,\!$ $^{174,}\,\!$ $^{178,}\,\!$ $^{180,}\,\!$ $^{188,}\,\!$ $^{214,}\,\!$ $^{216,}\,\!$ $^{218,}\,\!$ $^{244,}\,\!$ $^{246}$Hg. If the criterion is expanded to $|\Delta E| \le 1.2$ MeV, $^{182,184,186}$Hg are also candidates for shape coexistence.

In the experiment \cite{Bree:2014mxa, Rapisarda:2017hzb}, the energy differences of the ground states of $^{182,184,186}$Hg are about 0.4 MeV, while those from our calculations are about $1$ MeV. This show that the results of DRHBc are in  qualitative agreement with the experimental results for the $^{182,184,186}$Hg isotopes.

\subsection{Bubble structure with shape coexistence}\label{bcsc}

Now, we discuss a more exotic nucleus featuring both bubble structure and shape coexistence.
The experimental signatures due to exotic nuclear shapes are discussed, for examples, in Ref.~\cite{Garcia-Ramos:2018kcy} for shape coexistence and in ~\cite{Yong:2016zas, Otsuka:2018bqq, Fan:2019bcq, Sorlin:2020rwb} for bubble structures.

We expect that such interesting experimental features could correlate  in the case of a nucleus with both bubble structure and shape coexistence. For instance,
it was shown in Ref.~\cite{Yong:2016zas} the value of the $\pi^-/\pi^+$ ratio in the heavy ion
collision of bubble nuclei is larger than that in the collision of normal nuclei.
When we have bubble nuclei with shape coexistence in which only one of the almost degenerate
 vacua exhibits bubble structures,  the enhanced $\pi^-/\pi^+$ ratio due to bubble structures could be weakened.

\begin{table*}[b]
\begin{center}
\caption{List of the isotopes with both bubble structure and shape coexistence. $|\Delta E|~$ is the absolute value of the energy difference between  prolate and oblate shapes.
 \label{bubble_and_shape_coexistence}}
\vskip 0.3cm
\begin{tabular}{|c|c|c|c|c|c|}
  \hline
     \multirow{2}{*}{ Nucleus }             & \multicolumn{2}{c|}{ Prolate shape }  &  \multicolumn{2}{c|}{Oblate shape} & \multirow{2}{*}{$|\Delta E|$ [MeV] }        \\
     \cline{2-5}
      & $\beta_{2} $  & \phantom{xx} ${\cal B}_p^\star$ [$\%$] \phantom{xx}  & $\beta_{2}$ &  \phantom{xx}  ${\cal B}_p^\star$   [$\%$]  \phantom{xx} &  \\
  \hline
$^{202}$Hf & $+ 0.100$ & 23.7 & $ -0.072$  & 22.4 & 0.214 \\
\hline
$^{234}$Hf & $+0.272$ & 20.5 & $-0.234$ & 2.4 & 0.263 \\
\hline
$^{236}$Hf & $+0.262$ & 20.4 & $-0.227$ & 2.2 & 0.318 \\
\hline
$^{240}$Hf & $+0.241$ & 21.0 & $-0.192$ & 3.5 & 0.468 \\
\hline
$^{250}$Hf & $+0.091$ & 24.0 & $-0.086$ & 18.3 & 0.171 \\
  \hline
$^{194}$W & $+0.136$ & 23.1 & $-0.126$ & 12.1 & 0.155 \\
  \hline
$^{204}$W & $+0.080$ & 23.8 & $-0.060$ & 22.8 & 0.087 \\
  \hline
$^{254}$W & $+0.068$  & 24.2 & $-0.060$ & 21.7 & 0.401 \\
  \hline
$^{196}$Os & $+0.123$  & 23.6 & $-0.119$ & 11.0 & 0.096\\
  \hline
$^{206}$Os &  $+0.044$  & 24.1 & $-0.042$ & 23.5 & 0.006 \\
  \hline
$^{208}$Os & $+0.112$  & 22.4 & $-0.079$ & 20.4 & 0.447 \\
  \hline
$^{256}$Os & $+0.058$ & 24.0 & $-0.065$ & 20.6 & 0.141 \\
  \hline
$^{210}$Pt & $+0.072 $ & 22.5 & $-0.061$ & 21.4 & 0.016 \\
  \hline
$^{212}$Pt & $+0.113$ & 21.0 & $-0.080$ & 19.6 & 0.404  \\
  \hline
\end{tabular}
\end{center}
\end{table*}

We compile a list of the isotopes with both bubble structure and shape coexistence in Table \ref{bubble_and_shape_coexistence}. It can be seen from Table~\ref{bubble_and_shape_coexistence} that some isotopes such as $^{206}$Os have bubble structure  both in prolate and oblate shapes, while some of them, $^{196}$Os for example, possess it only in prolate shapes. This implies that the prolate shape supports bubble structure  more, which agrees with a general statement that the bubble parameter are tend to be smaller in oblate deformations than in prolate ones~\cite{Saxena:2018koy}.

\begin{figure}[t]
\centering
\subfigure{\includegraphics[width=.4\textwidth]{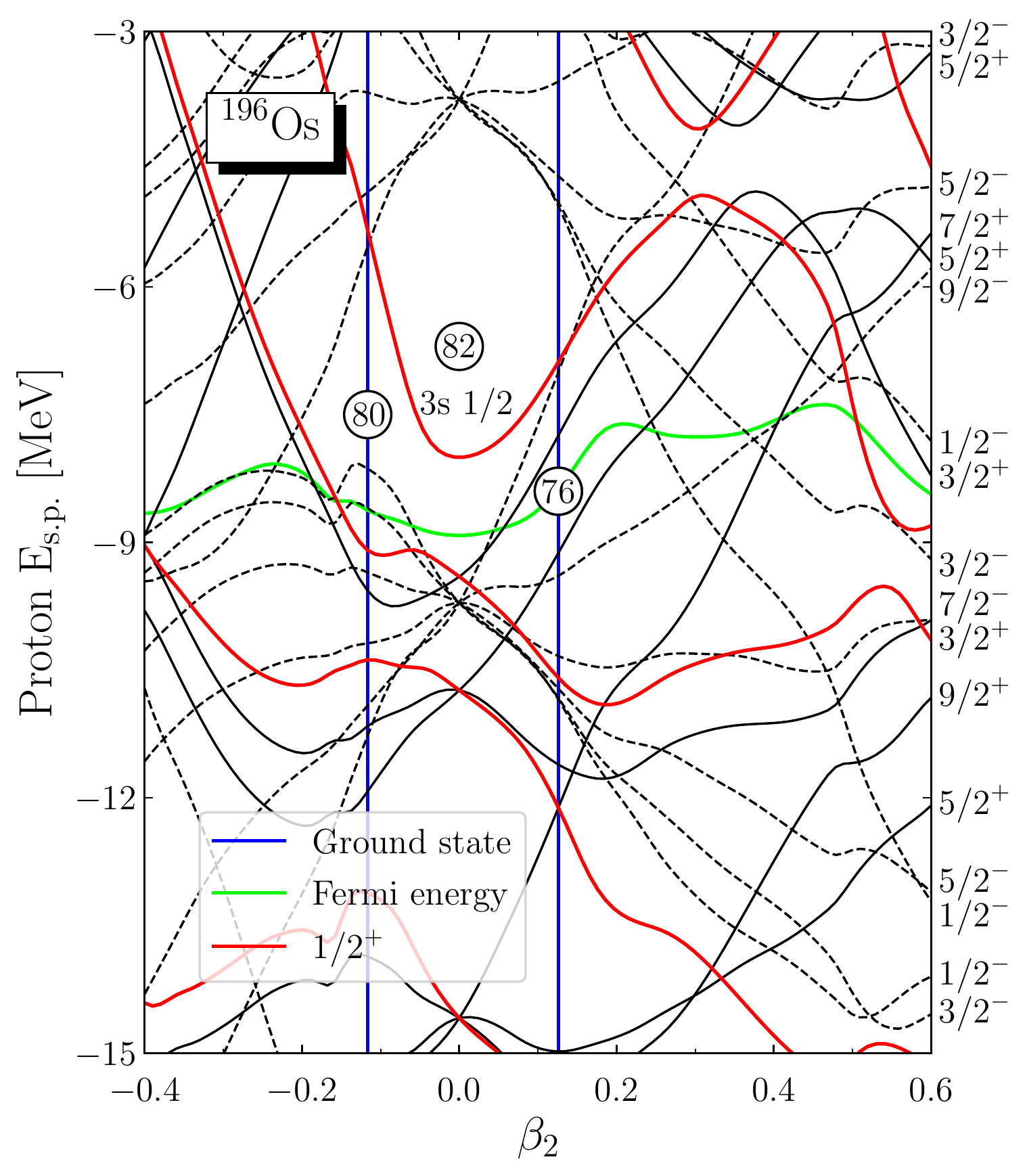}}
\subfigure{\includegraphics[width=.4\textwidth]{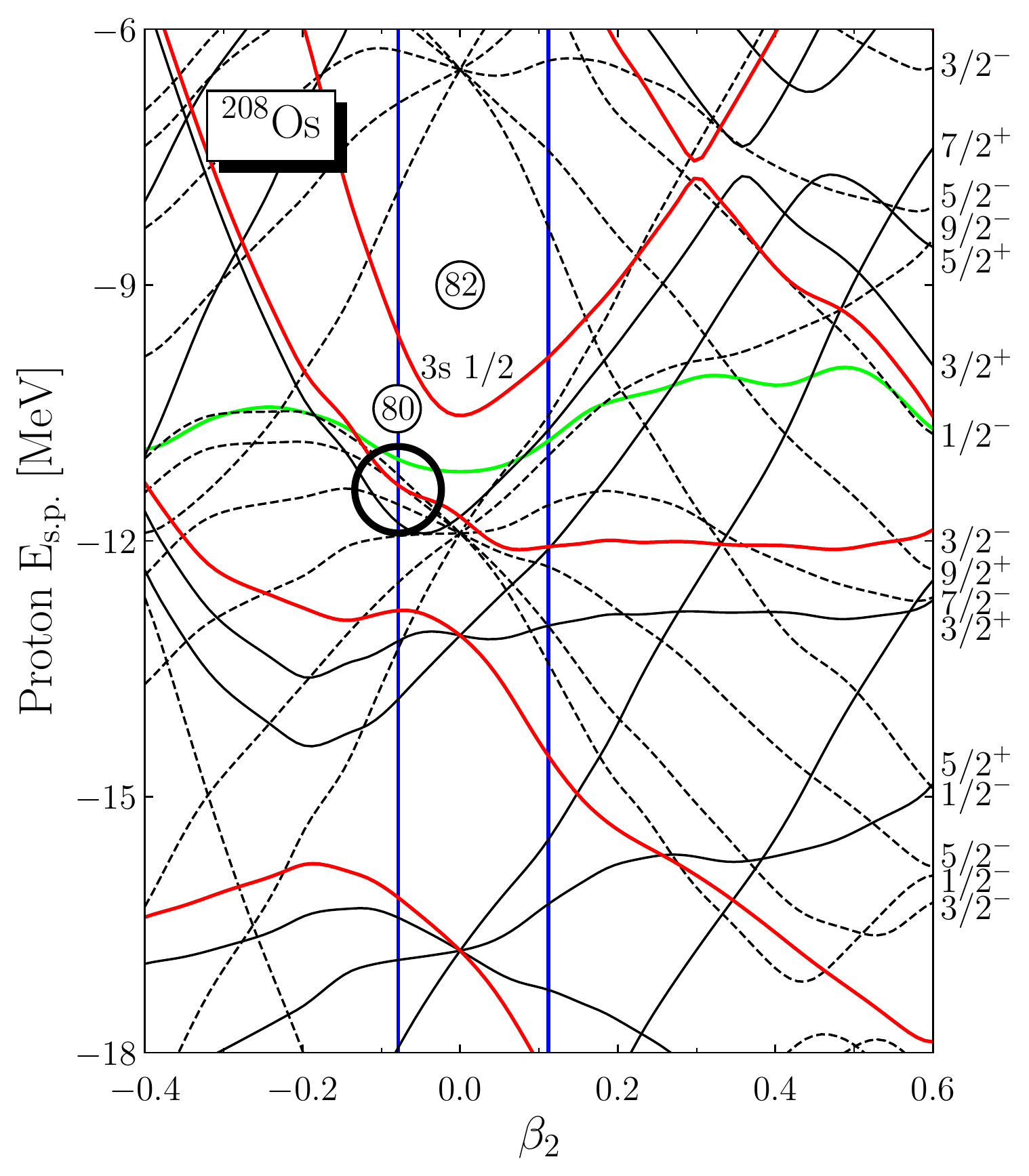}}
\caption{ The proton single-particle levels for $^{196}$Os and $^{208}$Os as a function of $\beta_2$.
Both isotopes exhibit shape coexistence. $^{196}$Os
has only prolate bubble, while $^{208}$Os possesses  prolate and oblate bubbles. The vertical solid lines indicate
the prolate or oblate minimum. Only $1/2^+$ states can contain s orbital directly related to central density. Near the Fermi surface, states are less occupied. Less occupied $1/2^+$ state near the Fermi surface can induce low central density and bubble structure. The gray lines represent the state except the $1/2+$ state. The solid lines and the dashed lines mean the positive and the negative parity state,  respectively. }
\label{Os_Nilsson_diagrams}
\end{figure}

\begin{figure*}[t]
\centering
\subfigure{\includegraphics[width=.4\textwidth]{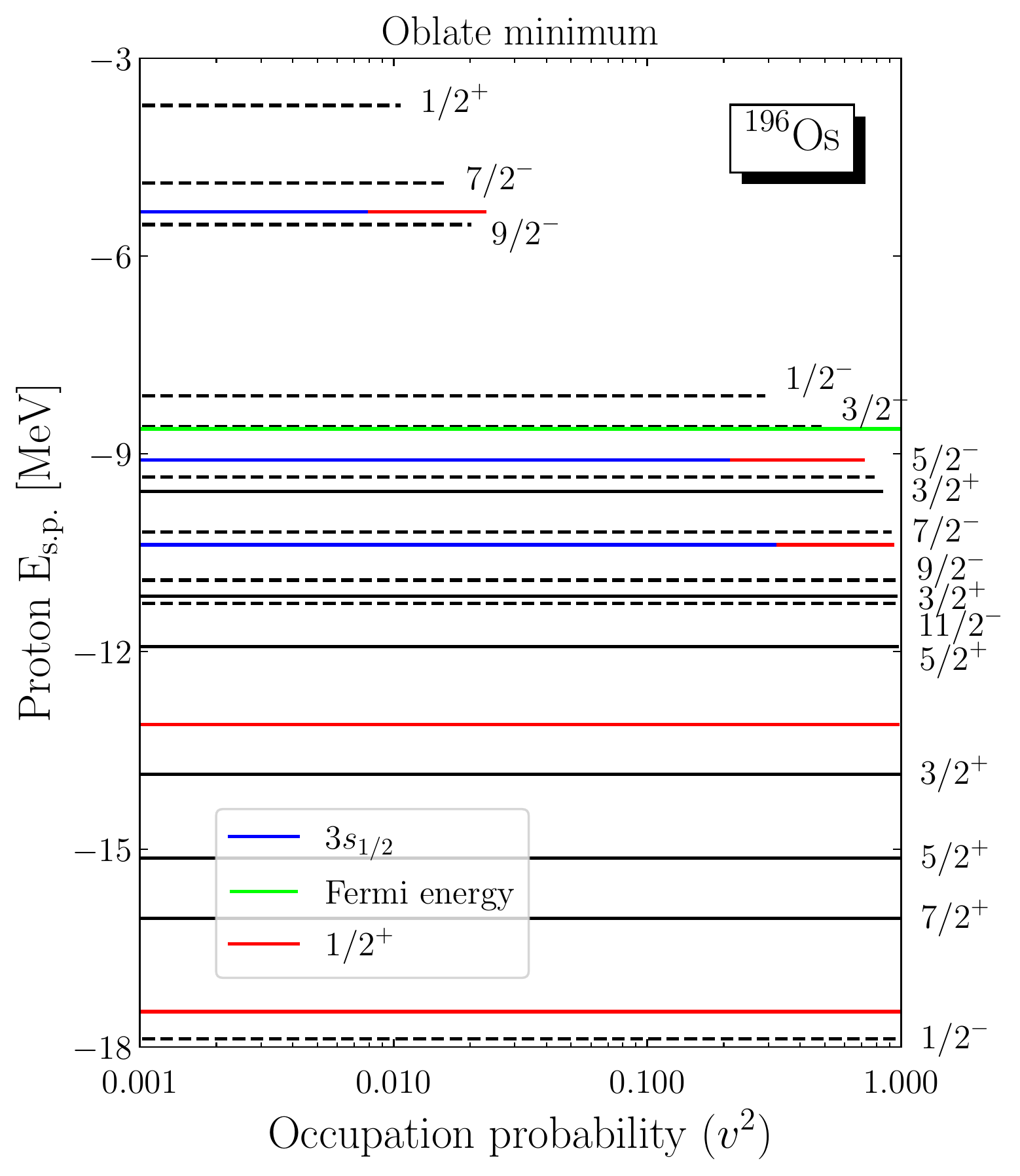}\label{Fig6a}}
\subfigure{\includegraphics[width=.4\textwidth]{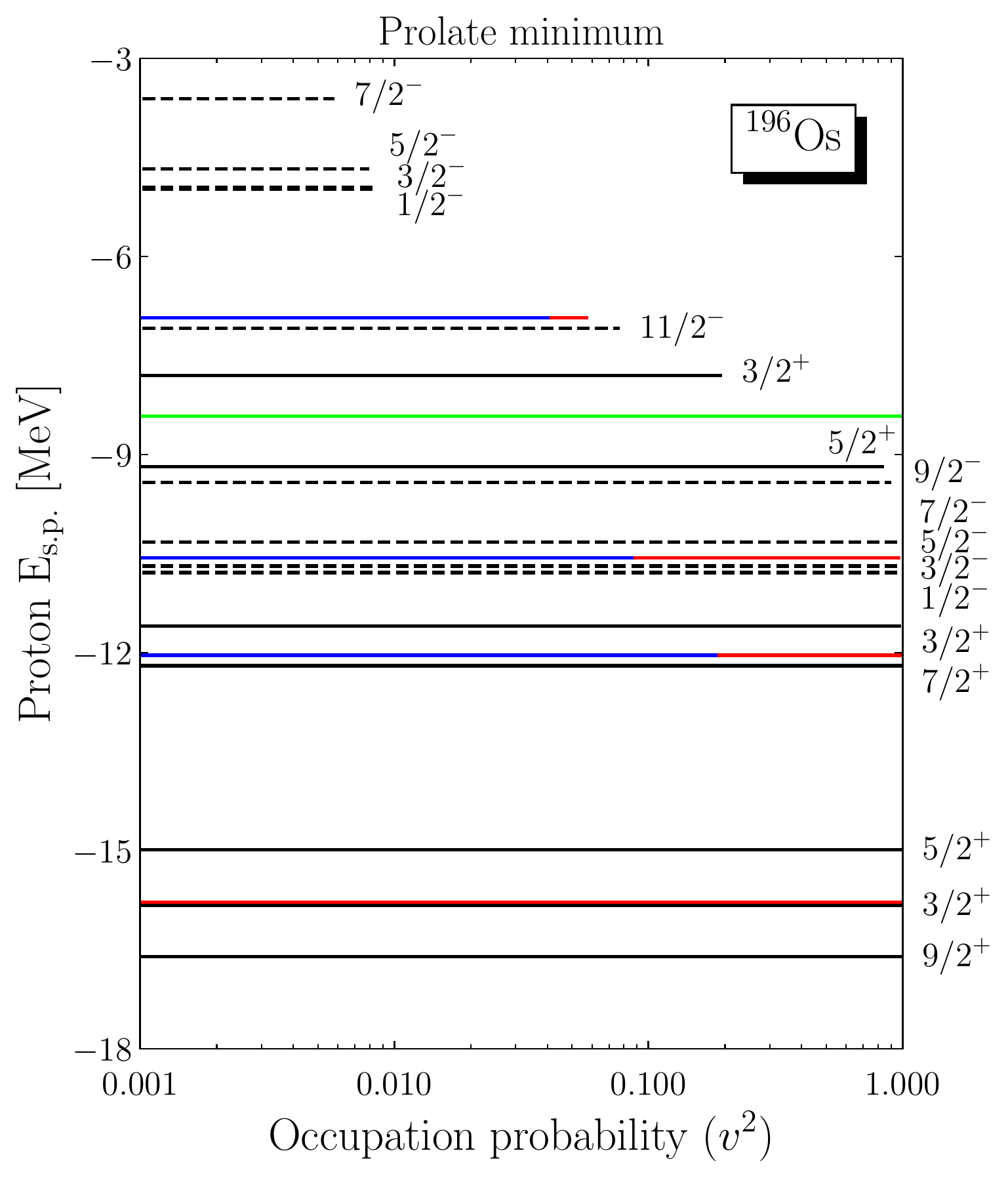}\label{Fig6b}}
\subfigure{\includegraphics[width=.4\textwidth]{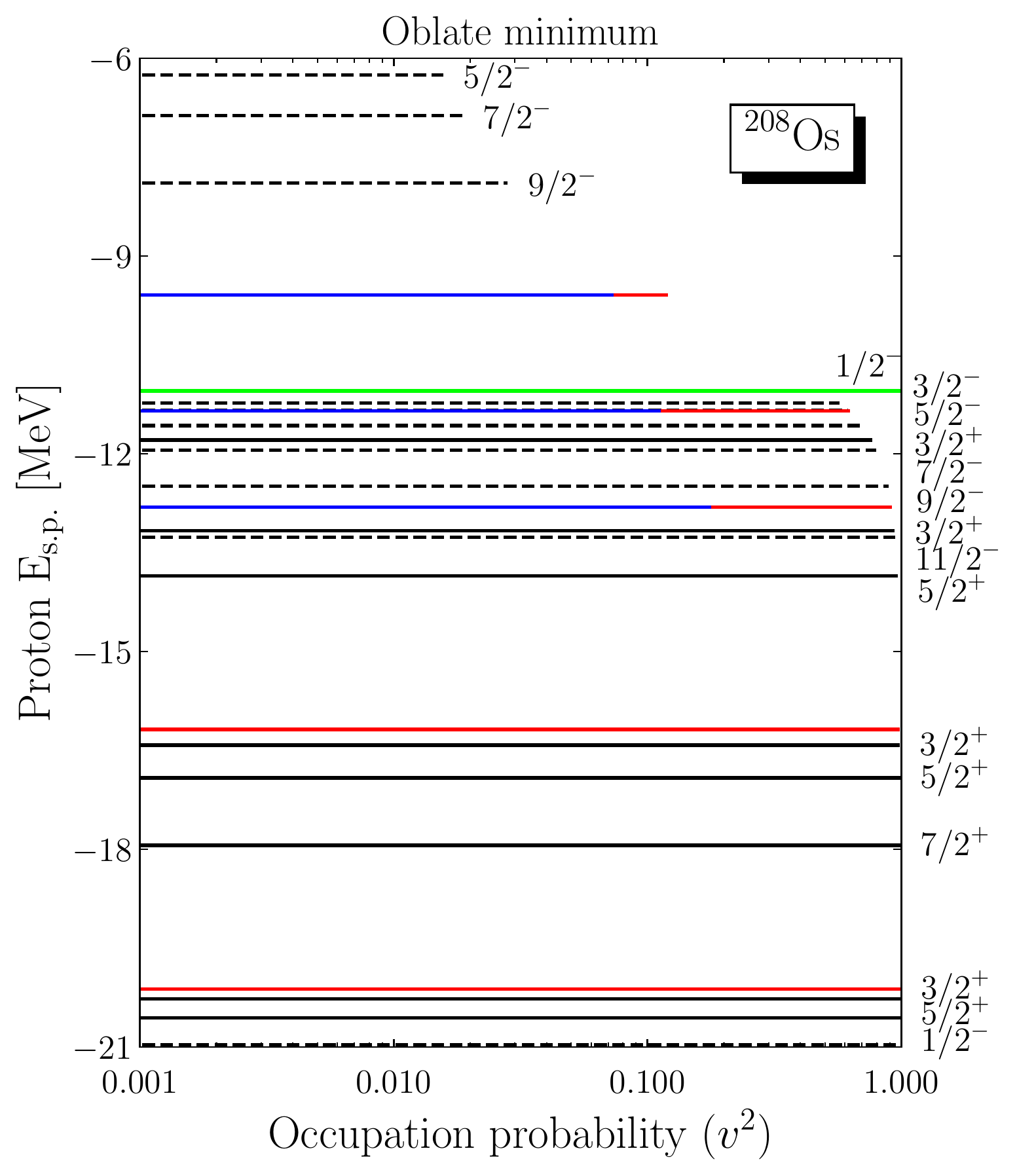}\label{Fig6c}}
\subfigure{\includegraphics[width=.4\textwidth]{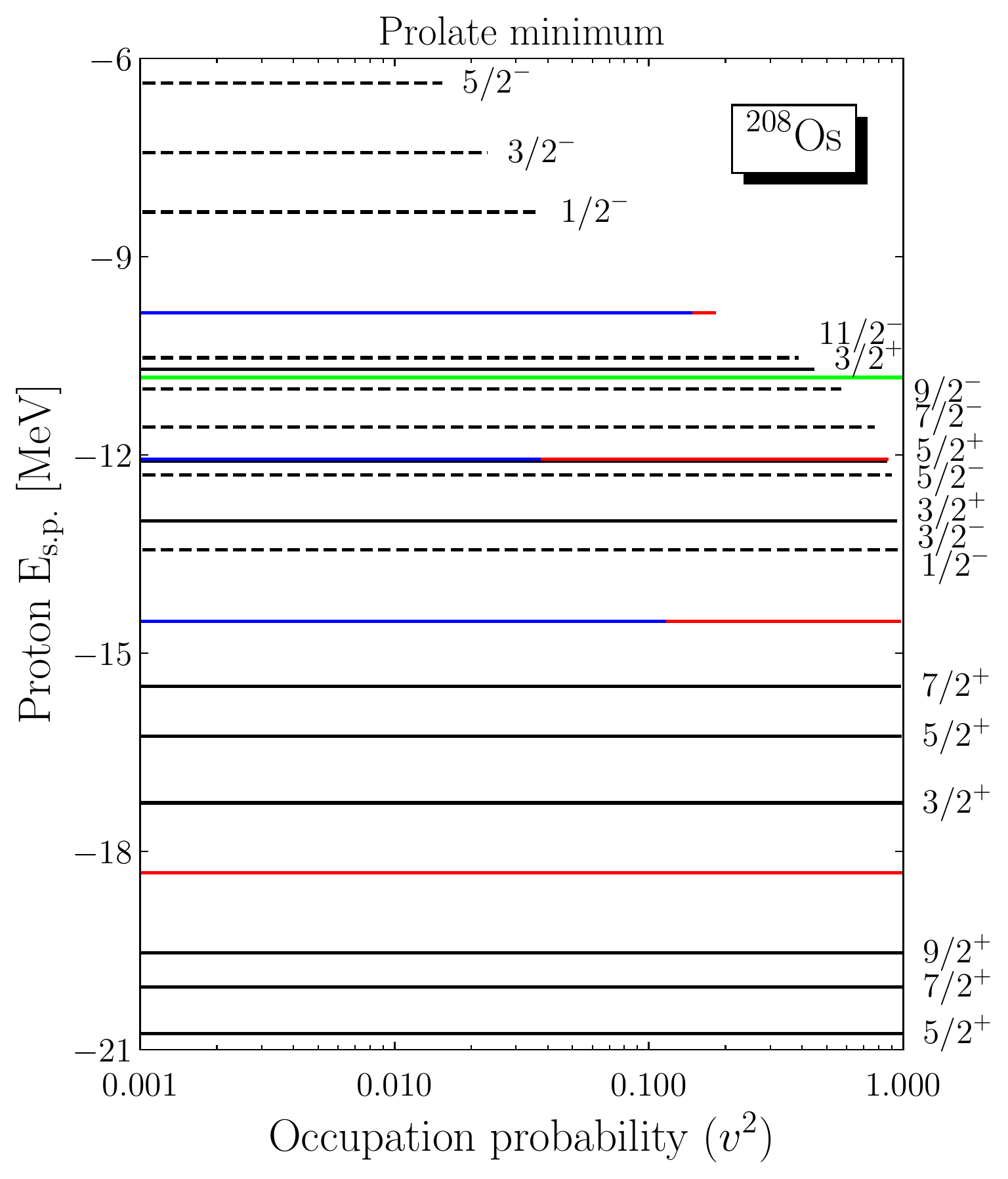}\label{Fig6d}}
\caption{ The proton single-particle levels around the Fermi energy for the ground states of $^{196}$Os and $^{208}$Os. All the 3s orbitals (blue solid lines) are hidden in $1/2^+$ states (red solid lines).  The 3s orbitals near the Fermi surface are related to low central proton density. It can be seen that 3s orbitals only exist in $1/2^+$ states. The grey lines represent the  states except the $1/2+$ state. The solid lines denote the positive parity state and dashed lines denote the negative parity state. }
\label{Os196_208_ground_states}
\end{figure*}

In Fig. \ref{Os_Nilsson_diagrams} we plot the proton single-particle levels for isotopes with two distinctive features: shape coexistence only with prolate bubble ($^{196}$Os) and  shape coexistence with both prolate and oblate bubble ($^{208}$Os). In prolate sides, the proton single-particle levels, especially $1/2^+$, for both isotopes show a similar trend leading to
prolate bubble structures in both cases, which is manifested from the occupation probabilities of the proton single-particle level in Fig. \ref{Os196_208_ground_states}. In Fig. \ref{Os196_208_ground_states}, 3s-state occupation amplitude in deformed nuclei is calculated by the following formula \cite{Sun:2021nyl}:
\be
N_{nlj}^{\rm DRHBc} = \langle \Psi | \hat{N}_{nlj} | \Psi \rangle = \langle \Psi | \sum_m c_{njlm}^\dagger c_{njlm} | \Psi \rangle \label{3s_state_occupation}
\ee
where $m$ denotes the total angular momentum projection on the symmetry axis. In oblate sides, the proton single-particle levels of $1/2^+$ for both isotopes may look similar, but as it is marked in black circle in Fig. \ref{Os_Nilsson_diagrams}, the two dashed levels close to the Fermi energy are quite different; in $^{208}$Os one is just above $1/2^+$, while the other level is below $1/2^+$, while in $^{196}$Os both levels are above $1/2^+$. Therefore, we can expect that  the $1/2^+$ level of $^{196}$Os is much occupied than that of  $^{208}$Os, leading to oblate bubble structure of $^{208}$Os as shown in Fig. \ref{Os196_208_ground_states}.

\section{Summary}\label{sec5}

We have studied the ground state properties of Hf, W, Os, Pt, and Hg even-even isotopes in the DRHBc framework, focusing on isotopes with both bubble structure and shape coexistence. For a systematic investigation, we first considered nuclear bubble structures and observed that deformations and pairing correlations hinder bubble structures by comparing
our results with those from RCHB.  Our results agree with previous studies \cite{Shukla:2014bsa, Luo:2018waj, Khan:2007ji, Grasso:2009zza, Saxena:2019ish, Yao:2012cx, Wang:2015coa} which also indicate that the deformation and the pairing correlation weaken the bubble structure.

After a brief study of shape coexistence, we searched for isotopes with bubble structure and shape coexistence and found candidate isotopes: $^{202}$Hf, $^{234}$Hf, $^{236}$Hf, $^{240}$Hf, $^{250}$Hf, $^{194}$W, $^{204}$W, $^{254}$W, $^{196}$Os, $^{206}$Os, $^{208}$Os, $^{256}$Os, $^{210}$Pt, and $^{212}$Pt. We analyzed the proton single-particle energy level for $^{196}$Os, shape coexistence only with prolate bubble, and  that for $^{208}$Os, shape coexistence with both prolate and oblate bubble. We expect that isotopes with bubble structure and shape coexistence may give distinctive experimental features. For instance, it was shown in Ref.~\cite{Yong:2016zas} the value of the $\pi^-/\pi^+$ ratio in the heavy ion collision of bubble nuclei is larger than that in the collision of normal nuclei. When we have bubble nuclei with shape coexistence in which only one of the almost degenerate vacua exhibits bubble structures, the enhanced $\pi^-/\pi^+$ ratio due to bubble structures could be weakened. It will be interesting to do simulations to quantify the importance of bubble nuclei with shape coexistence.
The beyond-mean-field effects have been implemented in DRHBc \cite{Sun:2021nfb} and the low-lying excited states of nuclei can be studied. These effects will influence the shape coexistence because the shape coexistence is directly related to the observed spectroscopy. 


\section*{ACKNOWLEDGMENTS}
The authors are grateful to the members of the DRHBc Mass Table Collaboration for useful discussions, to Gaurav Saxena for the discussion on the diffraction radius, and to X.W. Xia for providing the results of RCHB.
Part of this work has been developed during ``The 3$^{\rm rd}$ workshop on nuclear mass table with DRHBc theory'' supported by APCTP. The authors deeply appreciate the referee for reading our manuscript very carefully and giving very useful comments and suggestions.
YBC and CHL were supported by  National Research Foundation of Korea (NRF) grants funded by the
Korea government (Ministry of Science and ICT and Ministry of Education)  (No. 2016R1A5A1013277 and No. 2018R1D1A1B07048599).
This work was supported partly by the Rare Isotope Science Project of Institute for Basic Science funded by Ministry of Science, ICT and Future Planning and NRF of Korea (2013M7A1A1075764).
A portion of the computational resources were provided by the National Supercomputing Center  including technical support (KSC-2020-CRE-0329 and KSC-2021-CRE-0126).

\bibliographystyle{aip}

\end{document}